\documentclass{aastex62}
\newcommand{\ltsim}{\raisebox{-1.0ex}{$\stackrel{\textstyle<}{\sim}$}}
\def\kms{km~s$^{-1}$}
\def\al{Alfv\'{e}n}

\def\hinode{{\sl Hinode}}

\def\p78{{\sl P78-1}}

\def\soho{{\sl SOHO}}

\def\stereo{{\sl STEREO}}
\def\sdo{{\sl SDO}}

\def\caii{Ca~{\sc ii}}

\def\heii{He~{\sc ii}}

\def\halpha{H$\alpha$}

\def\al{Alfv\'{e}n}

\def\kms{km~s$^{-1}$}

\begin{document}

%

\title{Possible Production of Solar Spicules by Microfilament Eruptions}

\correspondingauthor{Alphonse C.~Sterling}
\email{alphonse.sterling@nasa.gov}

\author{Alphonse C. Sterling}
\affiliation{NASA/Marshall Space Flight Center, Huntsville, AL 35812, USA}

\author{Ronald L. Moore} 
\affiliation{NASA/Marshall Space Flight Center, Huntsville, AL 35812, USA}
\affiliation{Center for Space Plasma and Aeronomic Research, \\
University of Alabama in Huntsville, Huntsville, AL 35899, USA}

\author{Tanmoy Samanta}
\affiliation{Department of Physics and Astronomy, George Mason University, Fairfax, VA 22030, USA}
\affiliation{Johns Hopkins University Applied Physics Laboratory, Laurel, MD 20742, USA}

\author{Vasyl Yurchyshyn}
\affiliation{Big Bear Solar Observatory, New Jersey Institute of Technology, 40386 North Shore Lane,
Big Bear, CA 92314, USA}


\begin{abstract}

We examine Big Bear Solar Observatory (BBSO) Goode Solar Telescope (GST) high-spatial
resolution ($0''.06$), high-cadence (3.45~s), \halpha-0.8-\AA\ images of central-disk 
solar spicules, using data of \citet{samanta.et19}.  We compare with coronal-jet chromospheric-component observations of
\citet{sterling.et10a}.  Morphologically, bursts of spicules, referred to as ``enhanced
spicular activities" by \citet{samanta.et19}, appear as scaled-down versions of the jet's
chromospheric component. Both the jet and the enhanced spicular activities appear as chromospheric-material strands, 
undergoing twisting-type motions of $\sim$20---50~\kms\ in the jet and $\sim$20---30~\kms\ in the enhanced spicular activities.
Presumably, the jet resulted from a minifilament-carrying magnetic eruption. For two enhanced spicular activities
that we examine in detail, we find
tentative candidates 
for corresponding erupting {\it microfilaments}, but not expected 
corresponding base brightenings.  Nonetheless, the enhanced-spicular-activities' interacting mixed-polarity 
base fields, frequent-apparent-twisting motions, and morphological 
similarities to the coronal jet's chromospheric-temperature component,  
suggest that erupting microfilaments might drive the enhanced spicular activities but be hard to detect, perhaps due
to \halpha\ opacity.   Degrading the BBSO/GST-image resolution with a 
$1''.0$-FWHM smoothing function yields enhanced spicular activities resembling the ``classical spicules" described by, e.g., 
\citet{beckers68}.  Thus, a microfilament eruption might be the fundamental driver of many spicules, just as
a minifilament eruption is the fundamental driver of many coronal jets.  Similarly, a $0''.5$-FWHM smoothing 
renders some enhanced spicular activities to resemble previously-reported ``twinned" spicules, while
the full-resolution features might account for spicules sometimes appearing as 2D-sheet-like structures.


\end{abstract}


\section{Introduction}
\label{sec-introduction}

Solar spicules are deeply intriguing.  They shoot out from the chromosphere and reach $\sim$5$''$---10$''$
into the corona with a lifetime of a few minutes.  They have been observed for over 140 years
\citep{secchi1877} and  are omnipresent in the solar chromosphere, and yet we still lack a clear
understanding of what drives them.  The principle difficulty is that their widths are  $\ltsim$1$''$, and
hence at the limit of resolution of essentially all ground-based instruments  throughout the 19th and 20th
centuries.   Historically they were defined as features seen at the solar limb in chromospheric  emission
lines, such as \halpha\ or \caii, where they reach into the corona.  When looking at the chromosphere on the
solar disk,  features that almost certainly correspond to the limb spicules stem from the chromospheric
magnetic network.  

Much information regarding spicules had been  determined some time ago from  decades of ground-based observations, and their
properties are comprehensively summarized in several reviews 
\citep[e.g.,][]{beckers68,beckers72,bray.et74,michard74,zirin88,sterling00}.   More recent observations of spicules are from
both newer high-resolution ground-based imaging, and from seeing-free observations from space with the 2006-launched \hinode\
satellite using its own specific filter (3~\AA-wide \caii), but a one-to-one connection between spicules observed with those
earlier-era techniques and the newer observations has not been straightforward.  Henceforth, following the terminology
introduced in \citet{sterling.et10a} and \citet{pereira.et13}, we will use the term ``classical spicules" when referring to
observations and properties of spicules derived from the earlier-era observations, such as those described by the above-cited
pre-\hinode\ reviews. Because spicules are so numerous \citep[with estimates ranging from $\sim$10$^5$ to $10^6$ or more on
the Sun at any given time, e.g.,][]{athay59,beckers68,lynch.et73}, they have been suggested as possible contributors to
coronal heating \citep[][]{moore.et11,depontieu.et11,henriques.et16,samanta.et19}, although it is still unclear whether their
contribution to that heating is significant \citep[][]{madjarska.et11,klimchuk12,klimchuk.et14,bradshaw.et15}. Also,
mass-flux estimates of spicules indicate that if as little as 1\% of the apparently upward-moving  aggregate mass flux of
spicules escaped from the Sun, then spicules would supply the mass of the solar wind \citep[cf.,][]{tian.et14,samanta.et15}. 
For these reasons, understanding spicules is vital in considerations of the mass and energy  balance in the heliosphere. 
Other more-recent spicule observations from the ground include \citet{pasachoff.et09} and \citet{pereira.et16}.

Older ideas for spicule generation based on numerical simulations include energy inputs at
the  chromospheric base in the form of  single-pulse shocks \citep{suematsu.et82}, trains of
``rebound shocks" \citep{hollweg82}, torsional \al\ waves
\citep{hollweg.et82,kudoh.et99}; and energy releases in the middle or upper  chromosphere
\citep{sterling.et93}.  (See, e.g., the Beckers' reviews for ideas from the
pre-numerical-simulation era.)  None of these simulations however produced fully convincing
spicule models, especially given that the observations were generally of insufficient
quality to allow for an unambiguous characterization of spicule properties
\citep{sterling00}.

From the 21st century, new, higher-quality observations from the ground \citep[][]{rutten07}, and from space
\citep{depontieu.et04,depontieu.et07a}, as well as improved analysis techniques \citep{tavabi.et15}, and numerical
investigations \citep{iijima.et17,martinez-sykora.et17}, are
revolutionizing spicule studies.   Based largely on \hinode\ solar optical telescope (SOT) observations,
\citet{depontieu.et07a} argued that there are two types of spicules, which they called type~1 and type~2. (NB.
These designations are different from the type~1 and type~2 spicules defined in
\citeauthor{beckers68}~\citeyear{beckers68}; here we restrict our discussion only to type~1 and type~2 as defined by
\citeauthor{depontieu.et07a}~\citeyear{depontieu.et07a}.)  Type~1 spicules appear mainly in active region plage, 
and are relatively shorter-in-length
and more-slowly-moving spicules that tend to show both up and down motions clearly.  Type~2 spicules appear mainly in
quiet Sun and coronal holes, and are relatively fast moving and tend to show only upward motions clearly, with a
much fainter fall  \citep{skogsrud.et15}. \citet{depontieu.et07a} argue that type~1 spicules result from shocks in the
chromosphere \citep[but also see][]{shibata.et07}, and that type~2 spicules result from a different mechanism, likely
involving magnetic reconnection.  (See \citeauthor{zhang.et12}~\citeyear{zhang.et12} and
\citeauthor{pereira.et12}~\citeyear{pereira.et12} for lively discussions on whether there are two separate types
of spicules.)   Among reviews including some of the newer ideas for spicules are \citet{tsiropoula.et12},
\citet{zaqarashvili.et09}, and the subsection authored by T.~Pereira in \citet{hinode.et19}.
(\citeauthor{tsiropoula.et12} \citeyear{tsiropoula.et12}, and many other papers also, appear to assume that
type~1 spicules correspond to the historically observed classical spicules; as argued  by
\citeauthor{sterling.et10a}~\citeyear{sterling.et10a} and \citeauthor{pereira.et13}~\citeyear{pereira.et13}
however, if there are two different types, then the classical spicules correspond most closely to type~2
spicules, not type~1s.  Or it could be that classical observations saw both type~1 and type~2 spicules, 
according to T. Pereira in \citeauthor{hinode.et19}~\citeyear{hinode.et19}.)  Earlier, \citet{lee.et00} 
presented evidence that there is more than one type of spicule-size-scale chromospheric feature, but also pointed out
that the same fundamental feature might have different properties depending on its magnetic environment.

Many observations suggest that many spicules show spinning or twisting motions as they evolve.  Earlier, observations of
tilted spectral lines hinted at such spinning motions \citep[e.g.][]{beckers68,pasachoff.et68}. More recent
high-resolution observations from \hinode\ \citep{suematsu.et08} more strongly suggests twists, and
high-resolution ground-based spectral studies now confirm that at least some spicules twist \citep{depontieu.et12}.

Spicules -- or spicule-like features -- are also observed on the solar disk.  It might be said that the
classical versions of these are the ``mottles" of various types \citep[e.g.][]{beckers68,bray.et74}. 
More recent studies reveal new features that are suspected of being spicule counterparts, including
features called ``straws'' \cite{rutten07}, and ``Rapid blueshifted excursions'' (RBEs) and``rapid
redshifted  excursions'' (RREs) 
\citep[e.g.][]{langangen.et08,rouppe.et09,sekse.et12,sekse.et13a,sekse.et13b}.  These features can
display complex motions consisting of field-aligned flows, swaying motions, and also torsional
``spinning" motions \citep{sekse.et13b}.

Coronal jets also shoot out from the lower solar atmosphere, but they are larger than spicules and 
can reach $\sim$50{,}000~km \citep[e.g.,][]{shibata.et92,shimojo.et96,cirtain.et07,savcheva.et07}, and many of them
also show spin \citep[e.g.,][]{pike.et98,moore.et15}.  Most of them 
apparently are driven by eruptions of
minifilaments \citep{sterling.et15}, and most of these minifilament eruptions are seemingly prepared and 
triggered by canceling magnetic flux \citep{panesar.et16a}.  Motivated by suggestions by R. Moore
\citep{moore.et77,moore90}, \citet{sterling.et16a} postulated that spicules might be due to
eruptions of even-smaller-scale filaments that they called {\it microfilaments}.

Recently, \citet{samanta.et19} have obtained ten minutes of exceptional on-disk quiet Sun spicule observations
in \halpha\ ($\pm 0.8$~\AA) and high-resolution magnetograms from Big Bear Solar Observatory (BBSO)\@.  Those
observations revealed that bursts of spicule clumps, which they called ``enhanced spicular activities," were
apparently generated by interactions among mixed magnetic polarity elements at their bases.  In the present
paper, we take a second look at these enhanced spicular activities, and argue that they mimic, albeit on a
much smaller size scale, the larger-scale morphology and motions of chromospheric-temperature strands 
in a coronal jet.

\section{Coronal Jets in Chromospheric Lines}
\label{sec-jets}

We first address this question: What does a coronal jet look like when observed in chromospheric images?  In two cases, this
question has already been addressed.  Those studies  however were from 2010 \citep{sterling.et10a} and 2012
\citep{curdt.et12}, prior to our current understanding that many coronal jets result from minifilament eruptions prepared and
triggered by magnetic flux cancelation. With our new understanding, we reconsider what we are likely seeing in the
chromosphere when we look at coronal jets.

Figure~1 shows color-reversed \caii\ images of the solar limb from \hinode/SOT, showing
the same feature described in \citet{sterling.et10a}; this figure is similar to Figures~3
and~4 in that paper. A radial filter (due to T. E.
Berger and available as sot\_radial\_filter.pro in the SolarSoft software package) has been
applied; see
\citet{sterling.et10a} for further details. We have tuned the displayed intensities to highlight faint 
features in the images.  Nominally these \caii\ images show
spicules at the limb \citep[e.g.,][]{depontieu.et07a}.  From the Figure~1 images alone, it is not
fully apparent whether the features indicated by the white arrows in Figure~1(d) are unrelated
``type 2" spicules, or part
of the same $\sim$20$''$-wide structure.  Viewing the accompanying video to this figure
however suggests that they are indeed part of a single entity of this size, 
as its coherence is
maintained over 00:15:35---00:22:52~UT\@.  In fact, they are part of the same
{\it erupting} structure: \hinode/EIS 256~\AA\ \heii\ images clearly show that a corresponding broad feature erupts where
the Figure~1 feature occurs, and \hinode/XRT soft
X-ray images confirm that this feature coincides with an X-ray jet \citep{sterling.et10a}. Because
images such as those from \sdo/AIA were not available in 2007 (\sdo\ was launched in 2010), we could not
confirm whether the feature that underwent eruption was a minifilament, 
but its basic appearance and eruptive nature \citep[see Figs.~1(a) and~1(b) of][]{sterling.et10a} are 
consistent with it being similar to the erupting minifilaments frequently observed to make jets.   
In any case, those \hinode\ images confirm that {\it Figure~1 is showing the \caii-chromospheric
component of a X-ray coronal jet resulting from a small-scale eruption observed in EUV\@.}  Close 
inspection of video accompanying Figure~1 suggests that
the feature may be rotating in the \caii\ images, with the black arrows pointing to the left
in Figure~1(e---h) following a strand moving from left to right, and the black arrows
pointing to the right in Figure~1(6---i) following a feature moving right to left. This is
consistent with the jet manifesting as a partially transparent rotating cylinder viewed from the side,
in which visible strands stretch radially from the surface outward along the cylindrical jet spire,
with those strands nearest the observer carried in one direction and those in the
back side of the cylinder carried in the opposite direction.

We can make an estimate of the lateral velocity of the strands in the cylinder, projected against
the plane of the sky.  There is some uncertainty in identifying the same strands in different
images, and also we cannot be certain that some strands are not separate features crossing in front 
of or behind the erupting feature.  And, it turns out that we can find different velocities for 
some of the different strands.  If we consider the strand
pointed to by the black arrow in Figure~1(f), it moves about $8''$ over 00:18:51---00:20:53~UT,
which yields a speed of $\sim$50~\kms.  If instead however we consider the strand indicated by the
right-pointing arrow in Figures~1(g)---1(i), it only moves about $2''.5$ over 00:19:31---00:20:53~UT,
giving about 20~\kms. Other strands moving from left-to-right are closer to this lower velocity than the
above-derived 50~\kms; nonetheless, we will just say that the estimated rotational velocities are $\sim$20---50~\kms\
for this coronal jet.  If we assume that the feature on the left side, which is the 
fainter of the two, is on the far side of the cylinder, and that the feature on the right side is
in front, then the twisting motion would be clockwise when viewed from above.

As stated above, at the time of the \citet{sterling.et10a} paper, we had a much-less-complete understanding than now of
how most coronal jets work.  Much work by a number of workers have started to clarify the picture (for reviews
of jets, see \citeauthor{raouafi.et16}~\citeyear{raouafi.et16}, and the subsection authored by A. Sterling in
\citeauthor{hinode.et19}~\citeyear{hinode.et19}).  This is the probable scenario leading to the coronal jet
corresponding to the chromospheric strands of Figure~1:  Opposite-polarity magnetic flux elements converged at
the neutral line between the two polarities, over a  period lasting a couple of hours to as long as a couple of
days \citep{panesar.et17}. This resulted in formation of a magnetic flux rope, perhaps containing cool minifilament
material, that was rendered unstable by further convergence, and erupted to make the jet \citep{sterling.et15}.  
(Also see, e.g., \citeauthor{shen.et12}~\citeyear{shen.et12}, \citeauthor{huang.et12}~\citeyear{huang.et12}, \citeauthor{young.et14a}~\citeyear{young.et14a},  \citeauthor{hong.et14}~\citeyear{hong.et14}, \citeauthor{adams.et14}~\citeyear{adams.et14},   \citeauthor{wyper.et17}~\citeyear{wyper.et17}, \citeauthor{mcglasson.et19}~\citeyear{mcglasson.et19}.)  In \caii, 
apparently only the outlines of the erupting grossly cylindrical jet spire show up at locations where
strands of cool material are sufficiently dense (arrows in Fig.~1).

Furthermore, the twisting motions observed in many coronal jets 
\citep[e.g.][]{pike.et98,harrison.et01,patsourakos.et08,kamio.et10,raouafi.et10,curdt.et11,morton.et12,schmieder.et13,moore.et15}
plausibly might result when the minifilament field is twisted prior to eruption, and then during eruption transfers
its twist via reconnection to the ambient coronal field 
\citep{moore.et15,sterling.et15}, {\it a la} the dynamics proposed by \citet{shibata.et86}.  The
black arrows in Figure~1 might thus be showing the motions resulting from the untwisting
of twist injected into the system from the erupting-minifilament field.

\citet{curdt.et12} provide a second example of a coronal jet seen in \hinode/SOT \caii.  They
observed a polar coronal jet from 2007 November~4 in X-rays with \hinode/XRT and with  the
\hinode/EIS spectrometer, in addition to SOT\@.  They also observed the same jet with the
\soho/SUMER spectrometer and in EUV with \stereo/SECCHI, as described in
\citet{kamio.et10}. From XRT, this feature was clearly an X-ray coronal jet, and from
SECCHI it was  clearly a macrospicule jet in EUV 304~\AA\@.  \citet{curdt.et12} showed
that this coronal jet also had a clear chromospheric counterpart in SOT images, of width 
$\sim$20$''$, similar to the Figure~1 case.  
Moreover, the EIS and SUMER spectroscopic data showed Doppler evidence that
their jet was spinning. Thus, this is a second example of a coronal jet having a
strand-like chromospheric counterpart, and where the spire has a twisting (or untwisting) motion.

\section{High-resolution Spicules: Comparison with the Chromospheric Component of Coronal Jets}
\label{sec-spicules}

\citet{samanta.et19} observed quiet Sun spicules using ultra-high resolution adaptive
optics \halpha\ images from the 1.6m Goode Solar Telescope (GST) at Big Bear Solar
Observatory (BBSO)\@.  They observed at \halpha$\pm$0.8~\AA\ with a 0.07~\AA\ passband of
BBSO's VIS instrument, with a diffraction-limited resolution of $\sim$0$''.06$.  Their
ten-minute set of observations identified numerous bursts of clusters of narrow spicules, the enhanced spicular activities.  
In addition, there were a number of thin features they
identified as individual spicules in the data set.  They found that all 22
enhanced spicular activities that they observed with sufficient time coverage occurred in 
conjunction with rapidly evolving mixed-polarity magnetic
flux at their bases.  Thus they argued that the interactions of mixed-polarity magnetic
elements was crucial in the generation of the enhanced spicular activities.

First we will look at two of these enhanced spicular activities in detail.  
Figure~2 (and accompanying video) shows a closeup of the first one; this is the same
region as shown in panels A and D of movie S2 of \citet{samanta.et19}.  
This was presented as an example of an enhanced spicular activity resulting from a location of 
flux emergence in \citet{samanta.et19}, although we will reconsider this magnetic interpretation in 
\S\ref{sec-discussion}.  In these closeups, the enhanced spicular activity appears as 
a rotating cylinder, with the full width of the cylinder being approximately that between
the two white arrows in Figure~2(c), a distance covering approximately $1''$.  As with the
macrospicule of Figure~1, this apparent cylinder appears to have strands revealing rotation of
the cylinder with time.  For example, the strand indicated by the black arrows in panel~2(a) and 2(b) 
moves toward
the upper right over the time of these two panels, while the strand indicated by the blue arrow 
in Figures~2(a) and~2(b) moves
toward the lower left over the same period; this is analogous to the black arrows in Figure~1(e---i). 
To estimate the velocity of the possible rotations, we measure the speed of separation of 
the same two strands (pointed to by the black and blue arrows in Fig.~2(b)), over the time from
Figure~2(b) to Figure~2(c); the strands separate by $\sim$0$''.39$ over the 13.8~s between those two 
panels, which gives $\sim$21~\kms. 

Figure~3 (and accompanying video) shows our second detailed example from \citet{samanta.et19}; this 
is from their case that they argued resulted from magnetic flux cancelation.  Again, white arrows in
Figure~3(c) show the extent of the enhanced spicular activity.  In this case, spinning motion is less
apparent than in  the Figure~2 example. Instead, the dominant motion is a bodily shift of the entire 
structure upward (to the north) over, e.g., panels~3(b---c).  But, although somewhat  uncertain, there also
appears to be some twisting of the features, visible in the relative motions of the strands pointed to by
the  black and blue arrows in panels~3(b---c).  Over the times 4.54 to 4.66 min, the distance between the
two strands contracts (with the blue-arrow strand presumably rotating around the front of the black-arrow
strand) by $\sim$0.$''17$ over 7.2~s, giving $\sim$17~\kms. Thus the estimated spinning speeds of the
enhanced spicular activities in Figures~2 and~3 are about the same to within  the uncertainty of our
measurements, $\sim$20~\kms.  This is near the lower end of the range  of estimates for the twisting
velocity of the jet in Figure~1.

We have also inspected parameters of other enhanced spicular activities identified in \citet{samanta.et19}.
Figure~4(a) shows the full field of view of the \citet{samanta.et19}  study; this reproduces Figure~1(a) of
that paper, but this time with only the \halpha\ image, omitting the magnetic flux map. The accompanying
video (left panel) shows a time sequence of the  images over the entire $\sim$10~min observation
window.  Table~1 lists 22 enhanced spicular activities  from that video, giving the $(x,y)$ coordinates of
their locations (using the grid of Fig.~4), and their times as listed in the video.  These selected events
are the ones identified with  black circles in movie S4 of \citet{samanta.et19}.  In the table we give the
start and end time of each enhanced  activity, based on our observation of it in the movie.  The following
column gives the difference of  these times, which is the duration of the event.  We then indicate whether
we could observe twisting motion in the enhanced spicular activity, where `Y' means that there was
unambiguous apparent twisting motion (it is only ``apparent" since it is based on visual inspection only, as
we do not have Doppler data available).  We categorize four of the events as `W', indicating that we believe
that there is weak or short-duration spinning (perhaps only a small fraction of a rotation), but we are more
uncertain in these cases than in the `Y' cases as to whether there was actual spinning.  In some cases we
see splitting-type of motions of the enhanced spicular activities, similar to that described elsewhere
(e.g., \citeauthor{sterling.et10b}~\citeyear{sterling.et10b},
\citeauthor{yurchyshyn.et20}~\citeyear{yurchyshyn.et20}), where part of the feature splits off from the body
and expands away laterally; we denote this by `S' when we see splitting in the absence of obvious twisting
motions.  We denote cases without obvious spinning or splitting  with `N,' and cases that were uncertain
with `U.'  In cases where there was spinning and/or splitting (Y, S, W cases), we measured the velocity by
tracking features similar to what we did for the cases of Figures~2 and~3; the penultimate column of table~1
gives the time period over which we made these measurements, and the final column gives the determined
velocity along with an  uncertainty based on our estimates of the accuracy of our measurements.  The last
row of the table gives the averages for the duration and velocity.  

We find that about one-third of the events show clear visual evidence of twisting motions.  Four 
additional ones might show weak spinning, but this is somewhat uncertain.  On average the events have
durations of about three minutes, but it will be recalled that this is only for the spicular events of
sufficient velocity to appear in the passband defined by the GST filter centered on \halpha$\pm 0.8$~\AA;
therefore comparisons with other measurements of  durations of spicules/mottles must be made with caution.
Finally, the measured twisting/splitting  velocities (for the 14 cases where we could make estimates in
the final column) average about 30~\kms, but cover a wide range, as evidenced by the large sigma value of
15~\kms.  This velocity distribution however is bimodal, with the two spitting velocity measurements (64 and
58~\kms) skewing the average; with these two values removed, the mean of the remaining measured 12 
spinning and suspected-weak-spinning velocities is $22.5\pm 7.2$~\kms.

\section{Comparisons with ``Classical Spicules"}
\label{sec-classical spicules}

We can approximate the conditions of classical spicule observations, that is, the spicules
observed in chromospheric spectral lines from the ground with earlier-era methods 
\citep[e.g.,][]{beckers68,beckers72}.  Here we focus on the view of the BBSO/GST-observed
spicules with spatial resolution degraded to that similar to those of earlier studies.  Those
earlier studies additionally often had much poorer time cadence than the 3.45~s of our
GST data \citep{samanta.et19}.  Furthermore, the passband (transmission profile) of 
the \halpha\ filters used previously would have been of different quality than that of 
the Fabry-P{\'e}rot etalon of the \citet{samanta.et19} study.
Nonetheless, here we only consider consequences of degrading the spatial resolution
from that of the GST's $\sim$0$''.06$ resolution.  \citet{pereira.et13} did a similar degradation
of resolution of \hinode/SOT \caii\ limb spicules, and that added support to the idea that
the spicules identified as type~2 in the SOT images are the classical spicules.  Here we are repeating
the \citet{pereira.et13} exercise, but this time for the on-disk \halpha\ images we use here, 
in an effort to clarify the relationship between the features identified in \citet{samanta.et19} 
and the classical spicules.

The Figure~4(a) image, and accompanying video (left panel) shows the \citet{samanta.et19} region 
with full spatial resolution.  Figure~4(b) (and accompanying video, middle panel) shows the same image, but with a 
Gaussian smoothing of FWHM $0''.5$ applied
to each pixel in the image.  In the same fashion, Figure~4(c) (and accompanying video, right panel) shows the same image with
a Gaussian smoothing of FWHM $1''.0$ applied.  Accompanying videos show corresponding
time sequences of the images.

Compared to the full-resolution version, the FWHM $1''.0$ version in Figure~4(c) has lost 
a large amount of the structure of the enhanced spicular activities.  Indeed, when viewed as
a movie, these ``enhanced spicular activities" display group behavior, and are likely what would have been identified
as ``spicules," with widths of $\sim$1$''$, as reported by many of the earlier studies.  Thus
this figure and accompanying video show classical spicules (or 
``fine dark mottles").  

The middle-resolution version of FWHM $0''.5$ in Figure~4(b) shows the enhanced spicular activities to be
only  partially resolved.  In many cases only the edges of the enhanced spicular activities are prominent,
giving the spicules a double or twin structure, which has sometimes been reported
\citep{tanaka74,suematsu.et95,suematsu98}.

Returning again to the full-resolution version in Figure~4(a), the enhanced spicular activities are resolved into 
striations; this is suggestive of the ``2D sheet-like structure" for spicules, as described by
\citet{judge.et12}.  It is therefore understandable that the striations could 
appear as sheets under high resolution.  Different from \citet{judge.et12} however, we argue that 
many spicules could result from an upward mass flow of material, driven by an erupting microfilament 
undergoing reconnection with surrounding magnetic
field.  \citet{sekse.et13a} and \citet{pereira.et16} have already argued that the sudden
appearance of sheet-like spicules could be due to transverse and/or torsional motions
along the line of sight in the spicule body, and this explanation seems fully plausible.

As mentioned in the introduction, assuming that there are two spicule types, the enhanced spicular
activities of \citet{samanta.et19} would likely correspond to the type~2 spicules, and therefore our
findings of this section are in agreement with the conclusions of \citet{pereira.et13}. Our degradation
exercise here demonstrates that the on-disk features we observe here likely correspond to the limb spicules
of \citet{pereira.et13}, and it shows that the ``enhanced spicular activities" are in fact the classical
spicules/mottles, something that was not  immediately apparent in the \citet{samanta.et19} study. This
result might have been partially anticipated, given that \citet{pereira.et13} argued that type~II spicules
can be degraded to classical spicules, and that various works
\citep[e.g.][]{langangen.et08,rouppe.et09,sekse.et12,sekse.et13a,sekse.et13b} argue that RBEs/RREs are the
on-disk  representation of specifically type~II spicules. In some cases it was suggested that a single
larger RBE \citep[e.g.][]{sekse.et13b} or spicule \citep[e.g.][]{pereira.et16} might be composed  of thinner
strands.   Our work however emphatically emphasizes this last point: we argue that the clusters of strands
that make up enhanced spicular activities are plausibly part of a single larger-scale  spicule structure (often
of width similar to those given for classical spicules: $\sim$0$''$.5---1$''$.0), analogous to how the
strands that make up the chromospheric component of the coronal jet in Figure~1 are all part of the same
macroscopic EUV/X-ray coronal jet.

\section{Summary and Discussion} 
\label{sec-discussion}

Comparisons between high-resolution observations of spicules from BBSO/GST, and observations of the
chromospheric component to coronal jets, shows that it is plausible to consider that some population
of spicules could be scaled-down versions of coronal jets.  

If we assume for a moment that this is that case, that is, that many spicules are miniature versions of 
coronal jets, then we can then explain many of the GST high-resolution observations, and we can also 
offer explanations for several other previously observed spicule properties.  We make these arguments
in the following paragraphs.

Many coronal jets result from eruption of a minifilament.  It is - apparently - only some strands  of
the column (spire) of the jet that are visible in chromospheric images.  Thus in the case of Figure~1,
we see selections of cool-material strands that connect the photospheric magnetic flux with the
jet-spire field  that has undergone reconnection with an erupting minifilament flux rope; an
appropriate erupting feature, perhaps a cool-material minifilament, is  apparent in \hinode/EIS slot
images, and the resulting  soft X-ray jet spire is apparent in \hinode/XRT images
\citep{sterling.et10a}; similarly, under our assumption, spicules, such as those in Figures~2 and~3,
would be the chromospheric component -- visible at the observed GST wavelength -- of the spire of a miniature
jet-like feature made by an erupting microfilament flux rope.   Coronal jets show spin, presumably
because the erupting  minifilament's field is twisted and unleashes its twist onto neighboring open (or
far-reaching) field via interchange reconnection \citep{moore.et15}; spicules similarly often show twist
\citep[e.g.,][]{pasachoff.et68,depontieu.et14a}, and this could result from an erupting microfilament
field having twist that it imparts onto the spicule field via reconnection.  Moreover, the twist
velocities we estimate for the enhanced spicular activities, $\sim$20---30~\kms, match well 
with the estimated/observed spicule 
Doppler twist-speed values of $\sim$30~\kms\ of  \citet{pasachoff.et68} and 10---30~\kms\ of
\citet{depontieu.et14a}, and similar observed  transverse oscillational velocities \citep[see,
e.g.,][]{zaqarashvili.et09}, \hinode/SOT-observed limb-spicule horizontal speeds 
\citep{pereira.et12}, and some measurements of RBE transverse speeds \citep{sekse.et13a}.

The minifilament flux ropes that erupt to form coronal jets are
apparently built up by canceling opposite-polarity magnetic fields \citep{panesar.et17}; spicules also
apparently form at locations of interactions among opposite-polarity  fields \citep{samanta.et19}.
While for coronal jets the dominant mechanism that builds the minifilament field apparently is magnetic flux
cancelation \citep[e.g.,][]{panesar.et17,mcglasson.et19}, \citet{samanta.et19} on the other hand advocated
both magnetic cancelation and emergence as possible causes for spicules.  We caution however that the
fields involved with spicule creation might be quite weak, of order 10~G or weaker.  It is possible that
some of the apparently emerging fields identified by \citet{samanta.et19} at the base of spicules could
also have weak elements that are undergoing cancelation.  Moreover, it is common for coronal jets to
form at locations where the minority polarity pole of an emerging bipole cancels with surrounding majority
polarity field \citep[e.g.,][]{shen.et12,sterling.et17,panesar.et18a}.  Thus, the enhanced spicular activity events identified 
in \citet{samanta.et19} as possibly resulting from emergence episodes (such as our Fig.~2 event) might 
in fact be prepared and triggered by cancelation instead. Even-higher-resolution magnetograms
and spicule images will be required to determine whether emergence in the absence of cancelation sometimes 
produces spicules.   (\citeauthor{wang.et98}~\citeyear{wang.et98} found evidence that 
flux cancelation might be responsible for some high-speed spicules.)

If erupting microfilaments produce spicules, then there would be a natural scaling with three peaks in size 
range for filament-like eruptions causing transient solar  activity \citep{sterling.et16a}: (i)
active-region-scale filament eruptions  that drive solar eruptions and CMEs, (ii) supergranule-scale
minifilament eruptions  resulting in coronal jets,  and (iii) granule-scale microfilament eruptions that
result in spicules.  Thus, the three peaks in size scale for the  filament-like features that erupt would
reflect three natural magnetic size scales in the Sun: (i) active regions, (ii) supergranules, and (iii)
granules.  Moreover, magnetic flux cancelation might be the buildup and trigger mechanism for many of the
eruptions in all three cases, respectively for (i) large filament eruptions 
\citep{sterling.et18,chintzoglou.et19}, (ii) minifilament eruptions, and (iii) microfilament eruptions.

Figure~5 shows a schematic of the postulated mechanism for producing spicules via microfilament eruptions.
The setup is based on the event of Figure~2, with corresponding magnetic elements labeled in Figures~2(a)
and~5(a).   The ambient field leans toward the right because the strong part of the supergranule network
field resides northeast of the field of view of the spicule in Figure~2, which corresponds to the left in
Figure~5. Figure~5(b) shows the microfilament flux rope field developing, presumably as the negative- and
positive-polarity elements cancel at the magnetic neutral line.  If  those canceling fields are sheared,
they will generate a flux rope containing twist \citep{vanball.et89}.   Figure~5(c) shows runaway
reconnection taking place  below the erupting microfilament (flare-like, or ``internal" reconnection,
meaning internal to the erupting microfilament field) and between the erupting microfilament field and the
ambient field (interchange, or ``external" reconnection). These reconnections make new closed and open (or
far-reaching) field lines, indicated by the dashed lines in the figure. (Corresponding reconnections are
defined in the minifilament-eruption case for jets  in \citeauthor{sterling.et15}~\citeyear{sterling.et15}.)
Figure~5(d) shows when the reconnection has eaten into the microfilament flux rope, ejecting
chromospheric-material microfilament material along the open field, to make the chromospheric spicule.   The
external reconnection transfers/injects any twist built up in the microfilament field onto the spicule
field, resulting in the sometimes-observed spinning (untwisting) spicule motion.  This picture is analogous
to that for the minifilament-eruption mechanism for making the coronal jets \citep{sterling.et15}.  

From our discussion in \S\ref{sec-classical spicules}, what we are calling a ``spicule" in this description 
could appear as an enhanced spicular activity and/or spicule sheet, a twinned spicule, or a classical spicule, depending on the observation 
methods and circumstances.  Thus, assuming that spicules are small versions of coronal jets is also 
consistent with the these descriptions of spicule-scale chromospheric features.

We can understand why the medium-resolution images of Figure~4(b) show some
features to appear to be double/twin  spicules if a spicule is a cylindrical
column where only some strands of semi-opaque cooler material show up in
chromospheric lines, as we believe we are observing with the jet of Figure~1. 
If we assume that the semi-opaque strands are uniformly distributed around the
edge of an approximately cylindrical spicule, then blurring the view (from the
side-on perspective of Fig.~1) would tend to show more opaque material  near the
two sides of the cylinder, where the apparent density of those strands along the
line of sight is highest.  Thus the spicular geometry expected in the case that
it is a scaled-down coronal jet could explain the sometimes-reported twinned
appearance of spicules  (other models might also expect a cylindrical spicule
shape, in which case the same explanation could apply).  Furthermore, our
$\sim$1$''$-degraded-resolution images (Fig.~4(c)) show that the enhanced
spicular activities of \cite{samanta.et19} resemble what were previously
observed as on-disk classical spicules/mottles.  This agrees with the work first
done by \citet{pereira.et12} showing that some \hinode/SOT \caii\ limb spicules
resemble classical spicules.

Other observers have found stranded structure of spicules that appear to be similar to what we see, but at
the  limb \citep[e.g.,][]{pereira.et12,pereira.et16}.  A study by \citet{skogsrud.et14} found that it is
common for spicules to consist of multiple threads, with individual threads showing complex dynamics, with
groups of threads showing common behavior, and with some of them displaying torsional motions; this is fully
consistent with their observed features being the same as our enhanced spicular activities.   Other
high-resolution studies of on-disk spicules (RBEs/RREs) also show stranded structure
\citep[][]{yurchyshyn.et20}.  Some of the observations of finer, individual, spicules described by
\citet{samanta.et19} might be of substrands of an enhanced spicular activity, or themselves be composed of
still-finer strands.  Still however we cannot rule out that such individual spicules might have a different
driving mechanism all together.

Other expected characteristics of a microfilament-eruption model for spicules however are not obvious  in
the BBSO/GST spicule observations.  Many coronal jets are made by erupting minifilaments, and a 
brightening often appears to one side of the jet's base.  \citet{sterling.et15} called this a jet bright
point (JBP), and interpreted it as a small flare arcade occurring beneath the erupting minifilament, in 
analogy to a typical solar flare arcade occurring beneath erupting large-scale filaments.  If spicules
result from the same process, then we might expect evidence of an erupting microfilament and a
corresponding brightening; the small brown dashed loop at the spicule's base in Figures~5(c, d)
represents this brightening.  Also expected is brightening of a neighboring bipolar lobe at the
jet/spicule base; the larger dashed brown loop in Figures~5(c, d) represents this brightening. We do not
however see any hints of these brightenings in the BBSO/GST data.  \sdo/AIA~171~\AA\ movies of our
spicule region are provided in \citet{samanta.et19} (see movies S5---S9 in that paper); they show no
evidence for 171~\AA\  brightenings at the base in the enhanced spicular activities (classical
spicules/mottles).  We have inspected other AIA EUV channels, including 304, 193, and 211~\AA, and also
hotter-line AIA EUV channels, and we similarly find no evidence of base brightenings. As can been seen in
the 171~\AA\ movies of \citet{samanta.et19} however (and similarly for the other channels we have
inspected), both the spatial resolution and the time cadence of AIA are likely too low for a complete
assessment of this question.  AIA UV 1600 and 1700~\AA\ images show brightenings that closely match those
of the bright filigree pattern visible in our video vid4abc, but no obvious additional brightening, given
the resolution and cadence.  Thus the question of possible UV/EUV  spicule base brightenings will have to
be reconsidered with substantially higher resolution and cadence  instruments in the future.
(\citeauthor{panesar.et19}~\citeyear{panesar.et19} and 
\citeauthor{sterling.et20}~\citeyear{sterling.et20} find transient brightenings in high-resolution Hi-C
172~\AA\ EUV images  at the base of some small-scale features, but it is not yet known how those EUV jets
relate to typical spicules.)

More generally, we know of no clear, unambiguous such brightenings occurring near the time of spicule
initiation; \citet{suematsu.et95} reported brightenings that tend to begin at the time of peak spicule
extension or later, and so perhaps too late to correspond to the JBP\@.  \citet{sterling.et10b} reported
weak brightenings in \hinode/SOT \caii\ near-limb spicule images, which might be candidates for  JBP
counterparts if they can be confirmed.   While the lack of clear brightenings near the base of spicules in
the far wings of \halpha\ could be an argument that spicules are not due to microfilament eruptions, at the
same time we note that the JBP  is not apparent in the  chromospheric images of Figure~1, even though it is
visible in XRT X-ray images at comparable times  \citep[][Fig.~1(k,l)]{sterling.et10a}.  For that Figure~1
jet we do not know whether the erupting features (the suspected minifilament) would have been seen in \caii,
because the SOT observations for this event started  after the eruption visible in EIS had taken place.

A possibility however is that both an erupting microfilament and corresponding brightenings might not
be easily observable in high-core-opacity chromospheric spectral lines such as \halpha.  This is
because the putative erupting microfilament might might have a very slow velocity while it is still rising as
part of an in-tact flux rope, and the brightening features similarly might be stationary or moving very slowly.  In
that case, they  might not be visible in our line-wing images at \halpha-0.8~\AA\@. Also, such an
erupting microfilament and brightenings would be expected to form in the upper photosphere or low 
chromosphere, near where the  photospheric fields are canceling. Observations near the center of the
line, which would show low-velocity features, show emission from features higher in the chromosphere,
with lower-height features not visible  due to high opacity.  Therefore any potential erupting microfilaments
might be ``hiding" at wavelengths near the \halpha-line core!  In our data, we have identified a couple
of candidate absorption features that might be erupting microfilaments that are just beginning to move rapidly
enough to appear in the \halpha-0.8~\AA\ images; green arrows in the bottom panels of Figures~2(c)
and 3(a) show these features.  It would be valuable to look for such features with
high-resolution, high-cadence images at several locations in the \halpha\ line, together with
high-quality magnetograms.  Such a project might be attempted at the BBSO/GST, the Swedish Solar
Telescope, the New Vacuum Solar Telescope, or with the upcoming DKIST facility.

The \citet{samanta.et19} observations of spicules makes it hard to see how they could be driven by
shock waves  produced by photospheric motions, because mixed-polarity magnetic field seems to be
required to generate these spicules. Numerical simulations had been pointing to this conclusion for
some time  \citep[e.g.][]{sterling.et90,martinez-sykora.et13}.  (Whether such shocks drive the features
seen in plages called type~1 spicules (\citeauthor{depontieu.et07a}~\citeyear{depontieu.et07a}),  or
anemone jets (\citeauthor{shibata.et07}~\citeyear{shibata.et07}), is a different question, one that we
will not address here.)  Similarly, \al\ waves that are generated by photospheric motions alone would
not seem capable of producing the 20~\kms\ torsional motions that we observe in the low chromosphere
\citep{hollweg.et82,kudoh.et99}. If however the source of the torsional motions is something different
(perhaps an erupting twisted microfilament flux rope imparting its twist onto the spicule field, as
envisioned in Fig.~5), then a larger \al-wave amplitude might naturally result.  Inputing such
larger-amplitude \al\ waves at the base of the chromosphere might be a future direction for exploration
in \al-wave-driven spicule models \citep[e.g.][]{iijima.et17}.

Also, \citet{martinez-sykora.et17} were able to simulate so-called type~II spicules  by including  ambipolar
diffusion processes in a solar atmosphere model.  It appears that their model spicules  result when flux
tubes that are contorted by the high-beta plasma in and  below the photosphere spring from the photosphere
into the chromosphere.  Those contortions are released via ambipolar diffusion  of the field through the
plasma, and the resulting upward whipping motion of the field can eject material upward  as the spicule.  It
is still to be determined whether this proposed mechanism is consistent with  high-resolution spicule and
magnetic-field observations, such as those of \citet{samanta.et19}.  More generally, new observations may
support that more than one mechanism can produce what we call ``spicules."  From table~1, of the 22
enhanced spicular activities: seven show what we regard as clear visual evidence for spinning motions, four
are somewhat uncertain but might show weak spinning, another four show clear evidence for splitting, and
seven (including one of the splitting ones) are uncertain regarding whether they spin (meaning that we could
not determine with enough confidence to make a judgment).  Only one event (event 10) appeared to us not to
have either spinning or splitting behavior.   If only the spinning, weakly spinning, and splitting events
are due to minifilament eruptions, then about one-half of the enhanced spicular activities could be due to
something else.  Additionally, \citet{samanta.et19} identified numerous ``individual" or isolated spicules
\citep[probably equivalent to the ``isolated RBEs" of][]{sekse.et12}), and it is uncertain  whether
those isolated spicules are driven by the same mechanism as the ``enhanced spicular activity" spicules. 
Moreover, our observations are limited only to the subset of spicules that are prominent in our limited
observing range (\halpha$\pm 0.8$~\AA, corresponding to $\sim \pm$36~\kms), and therefore our observations
may omit a substantial portion of the spicule population.  Future observations should help  clarify whether
multiple mechanisms drive spicules.

\acknowledgments

A.C.S., T.S., and R.L.M. received funding from the  Heliophysics Division of NASA's Science Mission
Directorate  through the Heliophysics Guest Investigators (HGI) Program, and  the \hinode\ Project.
Hinode is a Japanese mission developed and launched by ISAS/JAXA, with NAOJ as domestic
partner and NASA and UKSA as international partners. It is operated by these agencies in
co-operation with ESA and NSC (Norway).

\bibliography{arxiv_spicule_200408}

\newcommand{\noop}[1]{}
\begin{thebibliography}{}
\expandafter\ifx\csname natexlab\endcsname\relax\def\natexlab#1{#1}\fi
\providecommand{\url}[1]{\href{#1}{#1}}
\providecommand{\dodoi}[1]{doi:~\href{http://doi.org/#1}{\nolinkurl{#1}}}
\providecommand{\doeprint}[1]{\href{http://ascl.net/#1}{\nolinkurl{http://ascl.net/#1}}}
\providecommand{\doarXiv}[1]{\href{https://arxiv.org/abs/#1}{\nolinkurl{https://arxiv.org/abs/#1}}}

\bibitem[{Adams {et~al.}(2014)Adams, Sterling, Moore, \& Gary}]{adams.et14}
Adams, M., Sterling, A.~C., Moore, R.~L., \& Gary, G.~A. 2014, Astrophysical
  Journal, 783, 11, \dodoi{10.1088/0004-637X/783/1/11}

\bibitem[{Athay(1959)}]{athay59}
Athay, R.~G. 1959, Astrophysical Journal, 129, 164, \dodoi{10.1086/146603}

\bibitem[{Beckers(1968)}]{beckers68}
Beckers, J.~M. 1968, Solar Physics, 3, 367, \dodoi{10.1007/BF00171614}

\bibitem[{Beckers(1972)}]{beckers72}
---. 1972, Annual Review of Astronomy and Astrophysics, 10, 73,
  \dodoi{10.1146/annurev.aa.10.090172.000445}

\bibitem[{Bradshaw \& Klimchuk(2015)}]{bradshaw.et15}
Bradshaw, S.~J., \& Klimchuk, J.~A. 2015, Astrophysical Journal, 811, 129,
  \dodoi{10.1088/0004-637X/811/2/129}

\bibitem[{Bray \& Loughhead(1974)}]{bray.et74}
Bray, R.~J., \& Loughhead, R.~E. 1974, The Solar Chromosphere (London: Chapman
  and Hall)

\bibitem[{Chintzoglou {et~al.}(2019)Chintzoglou, Zhang, Cheung, \&
  Kazachenko}]{chintzoglou.et19}
Chintzoglou, G., Zhang, J., Cheung, M. C.~M., \& Kazachenko, M. 2019, \apj,
  871, 67, \dodoi{10.3847/1538-4357/aaef30}

\bibitem[{Cirtain {et~al.}(2007)Cirtain, Golub, Lundquist, van Ballegooijen,
  Savcheva, Shimojo, DeLuca, Tsuneta, Sakao, Reeves, Weber, Kano, Narukage, \&
  Shibasaki}]{cirtain.et07}
Cirtain, J.~W., Golub, L., Lundquist, L., {et~al.} 2007, Science, 318, 1580,
  \dodoi{10.1126/science.1147050}

\bibitem[{Curdt \& Tian(2011)}]{curdt.et11}
Curdt, W., \& Tian, H. 2011, Astronomy and Astrophysics, 532, L9,
  \dodoi{https://doi.org/10.1051/0004-6361/201117116}

\bibitem[{Curdt {et~al.}(2012)Curdt, Tian, \& Kamio}]{curdt.et12}
Curdt, W., Tian, H., \& Kamio, S. 2012, Solar Physics, 280, 417,
  \dodoi{10.1007/s11207-012-9940-9}

\bibitem[{{De Pontieu} {et~al.}(2012){De Pontieu}, {Carlsson}, {Rouppe van der
  Voort}, {Rutten}, {Hansteen}, \& {Watanabe}}]{depontieu.et12}
{De Pontieu}, B., {Carlsson}, M., {Rouppe van der Voort}, L. H.~M., {et~al.}
  2012, Astrophysical Journal, 752L, 12, \dodoi{10.1088/2041-8205/752/1/L12}

\bibitem[{{De Pontieu} {et~al.}(2004){De Pontieu}, {Erd{\'e}lyi}, \&
  {James}}]{depontieu.et04}
{De Pontieu}, B., {Erd{\'e}lyi}, R., \& {James}, S.~P. 2004, Nature, 430, 536,
  \dodoi{10.1038/nature02749}

\bibitem[{{De Pontieu} {et~al.}(2011){De Pontieu}, {McIntosh}, {Carlsson},
  {Hansteen}, {Tarbell}, {Boerner}, {Martinez-Sykora}, \&
  {Title}}]{depontieu.et11}
{De Pontieu}, B., {McIntosh}, S.~W., {Carlsson}, M., {et~al.} 2011, Science,
  331, 55, \dodoi{10.1126/science.1197738}

\bibitem[{{De Pontieu} {et~al.}(2007){De Pontieu}, {McIntosh}, {Hansteen},
  {Carlsson}, {Schrijver}, {Tarbell}, {Title}, {Shine}, {Suematsu}, {Tsuneta},
  {Katsukawa}, {Ichimoto}, {Shimizu}, \& {Nagata}}]{depontieu.et07a}
{De Pontieu}, B., {McIntosh}, S., {Hansteen}, V.~H., {et~al.} 2007,
  Publications of the Astronomical Society of Japan, 59, 655,
  \dodoi{10.1093/pasj/59.sp3.S655}

\bibitem[{{De Pontieu} {et~al.}(2014){De Pontieu}, {Rouppe van der Voort},
  {McIntosh}, {Pereira}, {Carlsson}, {Hansteen}, {Skogsrud}, {Lemen}, {Title},
  {Boerner}, {Hurlburt}, {Tarbell}, {Wuelser}, \& {et al.}}]{depontieu.et14a}
{De Pontieu}, B., {Rouppe van der Voort}, L., {McIntosh}, S.~W., {et~al.} 2014,
  Science, 346, 1255732, \dodoi{10.1126/science.1255732}

\bibitem[{Harrison {et~al.}(2001)Harrison, Bryans, \& Bingham}]{harrison.et01}
Harrison, R.~A., Bryans, P., \& Bingham, R. 2001, Astronomy \& Astrophysics,
  379, 324, \dodoi{10.1051/0004-6361:20011171}

\bibitem[{Henriques {et~al.}(2016)Henriques, Kuridze, Mathioudakis, \&
  Keenan}]{henriques.et16}
Henriques, V. M.~J., Kuridze, D., Mathioudakis, M., \& Keenan, F.~P. 2016,
  Astrophysical Journal, 820, 124, \dodoi{10.3847/0004-637X/820/2/124}

\bibitem[{{Hinode Review Team} {et~al.}(2019){Hinode Review Team}, {Khalid},
  {Patrick}, {Baker}, R., \& {et al.}}]{hinode.et19}
{Hinode Review Team}, {Khalid}, A.-J., {Patrick}, A., {et~al.} 2019,
  Publications of the Astronomical Society of Japan, 71, id.R1,
  \dodoi{10.1093/pasj/psz084}

\bibitem[{Hollweg(1992)}]{hollweg82}
Hollweg, J.~V. 1992, Astrophysical Journal, 257, 345, \dodoi{10.1086/159993}

\bibitem[{Hollweg {et~al.}(1982)Hollweg, Jackson, \& Galloway}]{hollweg.et82}
Hollweg, J.~V., Jackson, S., \& Galloway, D. 1982, Solar Physics, 75, 35,
  \dodoi{10.1007/BF00153458}

\bibitem[{Hong {et~al.}(2014)Hong, Jiang, Yang, Bi, Li, Yang, \&
  Yang}]{hong.et14}
Hong, J., Jiang, Y., Yang, J., {et~al.} 2014, Astrophysical Journal, 796, 73,
  \dodoi{10.1088/0004-637X/796/2/73}

\bibitem[{Huang {et~al.}(2012)Huang, Madjarska, S., Doyle, \&
  Lamb}]{huang.et12}
Huang, Z., Madjarska, S., M., Doyle, J.~G., \& Lamb, D.~A. 2012, Astronomy and
  Astrophysics, 548, 62, \dodoi{10.1051/0004-6361/201220079}

\bibitem[{Iijima \& Yokoyama(2017)}]{iijima.et17}
Iijima, H., \& Yokoyama, T. 2017, Astrophysical Journal, 848, 38,
  \dodoi{10.3847/1538-4357/aa8ad1}

\bibitem[{Judge {et~al.}(2012)Judge, Reardon, \& Cauzzi}]{judge.et12}
Judge, P.~G., Reardon, K., \& Cauzzi, G. 2012, \apjl, 755, 11,
  \dodoi{10.1088/2041-8205/755/1/L11}

\bibitem[{Kamio {et~al.}(2010)Kamio, Curdt, Teriaca, Inhester, \&
  Solanki}]{kamio.et10}
Kamio, S., Curdt, W., Teriaca, L., Inhester, B., \& Solanki, S.~K. 2010,
  Astronomy and Astrophysics, 510, 1, \dodoi{10.1051/0004-6361/200913269}

\bibitem[{Klimchuk(2012)}]{klimchuk12}
Klimchuk, J.~A. 2012, Journal of Geophysical Research, 117, A12102,
  \dodoi{10.1029/2012JA018170}

\bibitem[{Klimchuk \& Bradshaw(2014)}]{klimchuk.et14}
Klimchuk, J.~A., \& Bradshaw, S.~J. 2014, Astrophysical Journal, 791, 60,
  \dodoi{10.1088/0004-637X/791/1/60}

\bibitem[{Kudoh \& Shibata(1999)}]{kudoh.et99}
Kudoh, T., \& Shibata, K. 1999, Astrophysical Journal, 514, 493,
  \dodoi{10.1086/306930}

\bibitem[{Langangen {et~al.}(2008)Langangen, De~Pontieu, Carlsson, Hansteen,
  Cauzzi, \& Reardon}]{langangen.et08}
Langangen, O., De~Pontieu, B., Carlsson, M., {et~al.} 2008, Astrophysical
  Journal, 679, 167L, \dodoi{10.1086/589442}

\bibitem[{Lee {et~al.}(2000)Lee, Chae, \& Wang}]{lee.et00}
Lee, C.-Y., Chae, J., \& Wang, H. 2000, Astrophysical Journal, 545, 1124,
  \dodoi{10.1086/317821}

\bibitem[{Lynch {et~al.}(1973)Lynch, Beckers, \& Dunn}]{lynch.et73}
Lynch, D.~K., Beckers, J.~M., \& Dunn, R.~B. 1973, Solar Physics, 30, 63L,
  \dodoi{10.1007/BF00156173}

\bibitem[{Madjarska {et~al.}(2011)Madjarska, Vanninathan, \&
  Doyle}]{madjarska.et11}
Madjarska, M.~S., Vanninathan, K., \& Doyle, J.~G. 2011, Astronomy and
  Astrophysics, 532, L1, \dodoi{10.1051/0004-6361/201116735}

\bibitem[{Mart{\'i}nez-Sykora {et~al.}(2017)Mart{\'i}nez-Sykora, De~Pontieu,
  Hansteen, Rouppe van~der Voort, Carlsson, \& Pereira}]{martinez-sykora.et17}
Mart{\'i}nez-Sykora, J., De~Pontieu, B., Hansteen, V.~H., {et~al.} 2017,
  Science, 356, 1269, \dodoi{10.1126/science.aah5412}

\bibitem[{Mart{\'i}nez-Sykora {et~al.}(2013)Mart{\'i}nez-Sykora, De~Pontieu,
  Leenaarts, Pereira, Carlsson, Hansteen, Stern, Tian, McIntosh, \& Rouppe
  van~der Voort}]{martinez-sykora.et13}
Mart{\'i}nez-Sykora, J., De~Pontieu, B., Leenaarts, J., {et~al.} 2013,
  Astrophysical Journal, 771, 66, \dodoi{10.1088/0004-637X/771/1/66}

\bibitem[{McGlasson {et~al.}(2019)McGlasson, Panesar, Sterling, \&
  Moore}]{mcglasson.et19}
McGlasson, R.~A., Panesar, N.~K., Sterling, A.~C., \& Moore, R.~L. 2019, \apj,
  882, 16, \dodoi{10.3847/1538-4357/ab2fe3}

\bibitem[{Michard(1974)}]{michard74}
Michard, R. 1974, in Proceedings from IAU Symposium no. 56 held at Surfer's
  Paradise, Qld., Australia, 3-7 September 1973. Edited by R. Grant Athay.
  International Astronomical Union. Symposium no. 56, ed. R.~G. Athay
  (Dordrecht; Boston: Reidel), 3

\bibitem[{Moore(1990)}]{moore90}
Moore, R.~L. 1990, Societ{\`a} Astronomica Italiana, Memorie (MmSAI), 61, 317

\bibitem[{Moore {et~al.}(2011)Moore, Sterling, Cirtain, \&
  Falconer}]{moore.et11}
Moore, R.~L., Sterling, A.~C., Cirtain, J.~W., \& Falconer, D.~A. 2011,
  Astrophysical Journal, 731L, 18, \dodoi{10.1088/2041-8205/731/1/L18}

\bibitem[{Moore {et~al.}(2015)Moore, Sterling, \& Falconer}]{moore.et15}
Moore, R.~L., Sterling, R.~L., \& Falconer, D.~A. 2015, Astrophysical Journal,
  806, 11, \dodoi{10.1088/0004-637X/806/1/11}

\bibitem[{Moore {et~al.}(1977)Moore, Tang, Bohlin, \& Golub}]{moore.et77}
Moore, R.~L., Tang, F., Bohlin, J.~D., \& Golub, L. 1977, Astrophysical
  Journal, 218, 286, \dodoi{10.1086/155681}

\bibitem[{Morton {et~al.}(2012)Morton, Srivastava, \&
  Erd{\'e}lyi}]{morton.et12}
Morton, R.~J., Srivastava, A.~K., \& Erd{\'e}lyi, R. 2012, Astronomy \&
  Astrophysics, 542, 70, \dodoi{10.1051/0004-6361/201117218}

\bibitem[{Panesar {et~al.}(2017)Panesar, Sterling, \& Moore}]{panesar.et17}
Panesar, N.~K., Sterling, A.~C., \& Moore, R.~L. 2017, Astrophysical Journal,
  844, 131, \dodoi{10.3847/1538-4357/aa7b77}

\bibitem[{Panesar {et~al.}(2018)Panesar, Sterling, \& Moore}]{panesar.et18a}
---. 2018, Astrophysical Journal, 853, 189, \dodoi{10.3847/1538-4357/aaa3e9}

\bibitem[{Panesar {et~al.}(2016)Panesar, Sterling, Moore, \&
  Chakrapani}]{panesar.et16a}
Panesar, N.~K., Sterling, A.~C., Moore, R.~L., \& Chakrapani, P. 2016,
  Astrophysical Journal, 832L, 7, \dodoi{10.3847/2041-8205/832/1/L7}

\bibitem[{Panesar {et~al.}(2019)Panesar, Sterling, Moore, Winebarger, Tiwari,
  Savage, Golub, Rachmeler, Kobayashi, Brooks, Cirtain, De~Pontieu, McKenzie,
  Morton, Peter, Testa, Walsh, \& Warren}]{panesar.et19}
Panesar, N.~K., Sterling, A.~C., Moore, R.~L., {et~al.} 2019, Astrophysical
  Journal, 887L, 8, \dodoi{10.3847/2041-8213/ab594a}

\bibitem[{Pasachoff {et~al.}(2009)Pasachoff, Jacobson, \&
  Sterling}]{pasachoff.et09}
Pasachoff, J.~M., Jacobson, W.~A., \& Sterling, A.~C. 2009, Solar Physics, 260,
  59, \dodoi{10.1007/s11207-009-9430-x}

\bibitem[{Pasachoff {et~al.}(1968)Pasachoff, Noyes, \&
  Beckers}]{pasachoff.et68}
Pasachoff, J.~M., Noyes, R.~W., \& Beckers, J.~M. 1968, Solar Physics, 5, 131,
  \dodoi{10.1007/BF00147962}

\bibitem[{Patsourakos {et~al.}(2008)Patsourakos, Pariat, Vourlidas, Antiochos,
  \& Wuelser}]{patsourakos.et08}
Patsourakos, S., Pariat, E., Vourlidas, A., Antiochos, S.~K., \& Wuelser, J.~P.
  2008, Astrophysical Journal, 680, L73, \dodoi{10.1086/589769}

\bibitem[{Pereira {et~al.}(2012)Pereira, De~Pontieu, \&
  Carlsson}]{pereira.et12}
Pereira, T. M.~D., De~Pontieu, B., \& Carlsson, M. 2012, Astrophysical Journal,
  759, 18, \dodoi{10.1088/0004-637X/759/1/18}

\bibitem[{Pereira {et~al.}(2013)Pereira, De~Pontieu, \&
  Carlsson}]{pereira.et13}
---. 2013, Astrophysical Journal, 764, 69, \dodoi{10.1088/0004-637X/764/1/69}

\bibitem[{Pereira {et~al.}(2016)Pereira, Rouppe~van~der Voot, \&
  Carlsson}]{pereira.et16}
Pereira, T. M.~D., Rouppe~van~der Voot, L., \& Carlsson, M. 2016, Astrophysical
  Journal, 824, 65, \dodoi{10.3847/0004-637X/824/2/65}

\bibitem[{Pike \& Mason(1998)}]{pike.et98}
Pike, C.~D., \& Mason, H.~E. 1998, Solar Physics, 182, 333,
  \dodoi{10.1023/A:1005065704108}

\bibitem[{Raouafi {et~al.}(2010)Raouafi, Georgoulis, Rust, \&
  Bernasconi}]{raouafi.et10}
Raouafi, N.-E., Georgoulis, M.~K., Rust, D.~M., \& Bernasconi, P.~N. 2010,
  Astrophysical Journal, 718, 981, \dodoi{10.1088/0004-637X/718/2/981}

\bibitem[{Raouafi {et~al.}(2016)Raouafi, Patsourakos, Pariat, Young, Sterling,
  Savcheva, Shimojo, Moreno-Insertis, DeVore, Archontis, T{\"o}r{\"o}k, Mason,
  Curdt, Meyer, Dalmasse, \& Matsui}]{raouafi.et16}
Raouafi, N.~E., Patsourakos, S., Pariat, E., {et~al.} 2016, Space Science
  Reviews, 201, 1, \dodoi{10.1007/s11214-016-0260-5}

\bibitem[{Rouppe van~der Voort {et~al.}(2009)Rouppe van~der Voort, Leenaarts,
  de~Pontieu, Carlsson, \& Vissers}]{rouppe.et09}
Rouppe van~der Voort, L., Leenaarts, J., de~Pontieu, B., Carlsson, M., \&
  Vissers, G. 2009, Astrophysical Journal, 705, 272,
  \dodoi{10.1088/0004-637X/705/1/272}

\bibitem[{Rutten(2007)}]{rutten07}
Rutten, R.~J. 2007, in The Physics of Chromospheric Plasmas ASP Conference
  Series, Proceedings of the conference held 9-13 October, 2006 at the
  University of Coimbra in Coimbra, Portugal.. San Francisco, ed. P.~Heinzel,
  I.~Dorotovi{\u c}, \& R.~J. Rutten (San Francisco: Astronomical Society of
  the Pacific), 27

\bibitem[{Samanta {et~al.}(2015)Samanta, Pant, \& Banerjee}]{samanta.et15}
Samanta, T., Pant, V., \& Banerjee, D. 2015, Astrophysical Journal, 815, L16,
  \dodoi{10.1088/2041-8205/815/1/L16}

\bibitem[{Samanta {et~al.}(2019)Samanta, Tian, Yurchyshyn, Peter, Cao,
  Sterling, Erd{'e}lyi, Ahn, Feng, Utz, Banerjee, \& Chen}]{samanta.et19}
Samanta, T., Tian, H., Yurchyshyn, V., {et~al.} 2019, Science, 366, 890,
  \dodoi{10.1126/science.aaw2796}

\bibitem[{Savcheva {et~al.}(2007)Savcheva, Cirtain, Deluca, Lundquist, Golub,
  Weber, Shimojo, Shibasaki, Sakao, Narukage, Tsuneta, \& Kano}]{savcheva.et07}
Savcheva, A., Cirtain, J., Deluca, E.~E., {et~al.} 2007, Publications of the
  Astronomical Society of Japan, 59, 771, \dodoi{10.1093/pasj/59.sp3.S771}

\bibitem[{Schmieder {et~al.}(2013)Schmieder, Guo, Moreno-Insertis, Aulanier,
  Yelles~Chaouche, Nishizuka, Harra, Thalmann, Vargas~Dominguez, \&
  Liu}]{schmieder.et13}
Schmieder, B., Guo, Y., Moreno-Insertis, F., {et~al.} 2013, Astronomy and
  Astrophysics, 559, A1, \dodoi{10.1051/0004-6361/201322181}

\bibitem[{Secchi(1877)}]{secchi1877}
Secchi, A. 1877, Le Soleil, Vol.~2 (Paris: Gauthier-Villars)

\bibitem[{Sekse {et~al.}(2012)Sekse, Rouppe van~der Voort, \&
  De~Pontieu}]{sekse.et12}
Sekse, D.~H., Rouppe van~der Voort, L., \& De~Pontieu, B. 2012, Astrophysical
  Journal, 752, 108, \dodoi{10.1088/0004-637X/752/2/108}

\bibitem[{Sekse {et~al.}(2013{\natexlab{a}})Sekse, Rouppe van~der Voort, \&
  De~Pontieu}]{sekse.et13a}
---. 2013{\natexlab{a}}, Astrophysical Journal, 764, 164,
  \dodoi{10.1088/0004-637X/764/2/164}

\bibitem[{Sekse {et~al.}(2013{\natexlab{b}})Sekse, Rouppe van~der Voort,
  De~Pontieu, \& Scullion}]{sekse.et13b}
Sekse, D.~H., Rouppe van~der Voort, L., De~Pontieu, B., \& Scullion, E.
  2013{\natexlab{b}}, Astrophysical Journal, 769, 44,
  \dodoi{10.1088/0004-637X/769/1/44}

\bibitem[{Shen {et~al.}(2012)Shen, Liu, Su, \& Deng}]{shen.et12}
Shen, Y., Liu, Y., Su, J., \& Deng, Y. 2012, Astrophysical Journal, 745, 164,
  \dodoi{10.1088/0004-637X/745/2/164}

\bibitem[{Shibata {et~al.}(1992)Shibata, Ishido, Acton, Strong, \&
  Uchida}]{shibata.et92}
Shibata, K., Ishido, Y., Acton, L.~W., Strong, K. T.and~Hirayama, T., \&
  Uchida, Y. 1992, PASJ, 2, 173

\bibitem[{Shibata \& Uchida(1986)}]{shibata.et86}
Shibata, K., \& Uchida, Y. 1986, Solar Physics, 178, 379

\bibitem[{Shibata {et~al.}(2007)Shibata, Nakamura, Matsumoto, Otsuji, Okamoto,
  Nishizuka, Kawate, Watanabe, Nagata, UeNo, Kitai, Nozawa, Tsuneta, Suematsu,
  Ichimoto, Shimizu, Katsukawa, Tarbell, Berger, Lites, Shine, \&
  Title}]{shibata.et07}
Shibata, K., Nakamura, T., Matsumoto, T., {et~al.} 2007, Science, 318, 1591,
  \dodoi{10.1126/science.1146708}

\bibitem[{Shimojo {et~al.}(1996)Shimojo, Hashimoto, Shibata, Hirayama, Hudson,
  \& Acton}]{shimojo.et96}
Shimojo, M., Hashimoto, S., Shibata, K., {et~al.} 1996, Publications of the
  Astronomical Society of Japan, 48, 123, \dodoi{10.1093/pasj/48.1.123}

\bibitem[{Skogsrud {et~al.}(2014)Skogsrud, Rouppe van~der Voort, \&
  De~Pointieu}]{skogsrud.et14}
Skogsrud, H., Rouppe van~der Voort, L., \& De~Pointieu, D. 2014, Astrophysical
  Journal, 795, 23, \dodoi{10.1088/2041-8205/795/1/L23}

\bibitem[{Skogsrud {et~al.}(2015)Skogsrud, Rouppe van~der Voort, De~Pointieu,
  \& Pereira}]{skogsrud.et15}
Skogsrud, H., Rouppe van~der Voort, L., De~Pointieu, D., \& Pereira, T. M.~D.
  2015, Astrophysical Journal, 806, 170, \dodoi{10.1088/0004-637X/806/2/170}

\bibitem[{Sterling(2000)}]{sterling00}
Sterling, A.~C. 2000, Solar Physics, 196, 79, \dodoi{10.1023/A:1005213923962}

\bibitem[{Sterling {et~al.}(2010{\natexlab{a}})Sterling, Harra, \&
  Moore}]{sterling.et10a}
Sterling, A.~C., Harra, L.~K., \& Moore, R.~L. 2010{\natexlab{a}},
  Astrophysical Journal, 722, 1644, \dodoi{10.1088/0004-637X/722/2/1644}

\bibitem[{Sterling \& Mariska(1990)}]{sterling.et90}
Sterling, A.~C., \& Mariska, J.~T. 1990, Astrophysical Journal, 349, 647,
  \dodoi{10.1086/168352}

\bibitem[{Sterling \& Moore(2016)}]{sterling.et16a}
Sterling, A.~C., \& Moore, R.~L. 2016, Astrophysical Journal, 828, L9,
  \dodoi{10.3847/2041-8205/828/1/L9}

\bibitem[{Sterling {et~al.}(2010{\natexlab{b}})Sterling, Moore, \&
  DeForest}]{sterling.et10b}
Sterling, A.~C., Moore, R.~L., \& DeForest, C.~E. 2010{\natexlab{b}},
  Astrophysical Journal, 714L, 1, \dodoi{10.1088/2041-8205/714/1/L1}

\bibitem[{Sterling {et~al.}(2015)Sterling, Moore, Falconer, \&
  Adams}]{sterling.et15}
Sterling, A.~C., Moore, R.~L., Falconer, D.~A., \& Adams, M. 2015, Nature, 523,
  437, \dodoi{10.1038/nature14556}

\bibitem[{Sterling {et~al.}(2017)Sterling, Moore, Falconer, Panesar, \&
  Martinez}]{sterling.et17}
Sterling, A.~C., Moore, R.~L., Falconer, D.~A., Panesar, N.~K., \& Martinez, F.
  2017, Astrophysical Journal, 844, 28, \dodoi{10.3847/1538-4357/aa7945}

\bibitem[{Sterling {et~al.}(2018)Sterling, Moore, \& Panesar}]{sterling.et18}
Sterling, A.~C., Moore, R.~L., \& Panesar, N.~K. 2018, Astrophysical Journal,
  864, 68, \dodoi{10.3847/1538-4357/aad550}

\bibitem[{Sterling {et~al.}(2020)Sterling, Moore, Panesar, Reardon, Molnar,
  Rachmeler, Savage, \& Winebarger}]{sterling.et20}
Sterling, A.~C., Moore, R.~L., Panesar, N.~K., {et~al.} 2020, Astrophysical
  Journal, 889, 187, \dodoi{10.3847/1538-4357/ab5dcc}

\bibitem[{Sterling {et~al.}(1993)Sterling, Shibata, \& Mariska}]{sterling.et93}
Sterling, A.~C., Shibata, K., \& Mariska, J.~T. 1993, Astrophysical Journal,
  407, 778, \dodoi{10.1086/172559}

\bibitem[{Suematsu(1998)}]{suematsu98}
Suematsu, Y. 1998, in Solar Jets and Coronal Plumes, Proceedings of an
  International meeting, Guadeloupe, France, 23-26 February 1998, ESA, Vol.~42,
  ed. T.-D. Guyenne (Paris: European Space Agency (ESA)), 19

\bibitem[{Suematsu {et~al.}(2008)Suematsu, Ichimoto, Katsukawa, Shimizu,
  Okamoto, Tsuneta, Tarbell, \& Shine}]{suematsu.et08}
Suematsu, Y., Ichimoto, K., Katsukawa, Y., {et~al.} 2008, in First Results From
  Hinode ASP Conference Series, Vol. 397, Proceedings of the conference held
  20-24 August, 2007, at Trinity College Dublin, Dublin, Ireland., ed. S.~A.
  Matthews, J.~M. Davis, \& L.~K. Harra (San Francisco: Astronomical Society of
  the Pacific), 27

\bibitem[{Suematsu {et~al.}(1982)Suematsu, Shibata, Nishikawa, \&
  Kitai}]{suematsu.et82}
Suematsu, Y., Shibata, K., Nishikawa, T., \& Kitai, R. 1982, Solar Physics, 75,
  99, \dodoi{10.1007/BF00153464}

\bibitem[{Suematsu {et~al.}(1995)Suematsu, Wang, \& Zirin}]{suematsu.et95}
Suematsu, Y., Wang, H., \& Zirin, H. 1995, Astrophysical Journal, 450, 411,
  \dodoi{10.1086/176151}

\bibitem[{Tanaka(1974)}]{tanaka74}
Tanaka, K. 1974, in Chromospheric Fine Structure: Proceedings from IAU
  Symposium no. 56 held at Surfer's Paradise, Qld., Australia, 3-7 September
  1973., ed. R.~G. Athay (Dordrecht; Boston: International Astronomical Union),
  239

\bibitem[{{Tavabi} {et~al.}(2015){Tavabi}, {Koutchmy}, {Ajabshirizadeh},
  {Ahangarzadeh Maralani}, \& {Zeighami}}]{tavabi.et15}
{Tavabi}, E., {Koutchmy}, S., {Ajabshirizadeh}, A., {Ahangarzadeh Maralani},
  A.~R., \& {Zeighami}, S. 2015, Astronomy \& Astrophysics, 573, 7,
  \dodoi{10.1051/0004-6361/201423385}

\bibitem[{Tian {et~al.}(2014)Tian, DeLuca, Cranmer, De~Pontieu, Peter,
  Martínez-Sykora, Golub, McKillop, Reeves, Miralles, McCauley, Saar, Testa,
  Weber, Murphy, Lemen, Title, Boerner, Hurlburt, Tarbell, Wuelser, Kleint,
  Kankelborg, Jaeggli, Carlsson, Hansteen, \& McIntosh}]{tian.et14}
Tian, H., DeLuca, E.~E., Cranmer, S.~R., {et~al.} 2014, Science, 346,
  \dodoi{10.1126/science.1255711}

\bibitem[{Tsiropoula {et~al.}(2012)Tsiropoula, Tziotziou, Kontogiannis,
  Madjarska, Doyle, \& Suematsu}]{tsiropoula.et12}
Tsiropoula, G., Tziotziou, K., Kontogiannis, I., {et~al.} 2012, Space Science
  Reviews, 169, 181, \dodoi{10.1007/s11214-012-9920-2}

\bibitem[{van Ballegooijen \& Martens(1989)}]{vanball.et89}
van Ballegooijen, A.~A., \& Martens, P. C.~H. 1989, Astrophysical Journal, 343,
  971, \dodoi{10.1086/167766}

\bibitem[{Wang {et~al.}(1998)Wang, Johannesson, Stage, Lee, \&
  Zirin}]{wang.et98}
Wang, H., Johannesson, A., Stage, M., Lee, C., \& Zirin, H. 1998, Solar
  Physics, 178, 55, \dodoi{10.1038/nature22050}

\bibitem[{Wyper {et~al.}(2017)Wyper, Antiochos, \& DeVore}]{wyper.et17}
Wyper, P.~F., Antiochos, S.~K., \& DeVore, C.~R. 2017, Nature, 544, 452,
  \dodoi{10.1038/nature22050}

\bibitem[{Young \& Muglach(2014)}]{young.et14a}
Young, P.~R., \& Muglach, K. 2014, Solar Physics, 289, 3313,
  \dodoi{10.1007/s11207-014-0484-z}

\bibitem[{Yurchyshyn {et~al.}(2020)Yurchyshyn, Cao, Abramenko, Yang, \&
  Cho}]{yurchyshyn.et20}
Yurchyshyn, V., Cao, W., Abramenko, V., Yang, X., \& Cho, K.-S. 2020,
  Astrophysical Journal Letters, 891, 21, \dodoi{10.3847/2041-8213/ab7931}

\bibitem[{Zaqarashvili \& Erdl{\'e}yi(2009)}]{zaqarashvili.et09}
Zaqarashvili, T.~V., \& Erdl{\'e}yi, R. 2009, Solar Physics, 149, 355,
  \dodoi{10.1007/s11214-009-9549-y}

\bibitem[{Zhang {et~al.}(2012)Zhang, Shibata, Wang, Mao, Matsumoto, Liu, \&
  Su}]{zhang.et12}
Zhang, Y.~Z., Shibata, K., Wang, J.~X., {et~al.} 2012, Astrophysical Journal,
  750, 16, \dodoi{10.1088/0004-637X/750/1/16}

\bibitem[{Zirin(1988)}]{zirin88}
Zirin, H. 1988, Astrophysics of the Sun (Cambridge and New York: Cambridge
  University Press)

\end{thebibliography}

\clearpage


\begin{deluxetable}{lccclcc}
\tabletypesize{\footnotesize}
\tablecaption{``Enhanced Spicular Activity" Properties \label{tab:table1}}
\tablehead{
\colhead{Event} & \colhead{Location\tablenotemark{a} [($x,y$)]} & \colhead{Time Period\tablenotemark{b}} & \colhead{Duration [min]}& Twist?\tablenotemark{c} &\colhead{Measurement Period\tablenotemark{d}} & \colhead{Velocity\tablenotemark{e} [\kms]}
}
\startdata
1 & (14,19) & --- --- 2.42 & --- & U\tablenotemark{f} & --- & --- \\ 
2 & (22,23) & --- --- 2.24 & ---& Y & 1.26 --- 1.96 & $14\pm 2$ \\
3 & (28,18) & 0.34 --- 2.36 & 2.02 & Y & 1.78 --- 2.24 & $30\pm 3$ \\
4 & (19,14) & 0.17 --- 2.59 & 2.42 & S & 0.75 --- 1.44 & $64\pm 10$ \\
5 & (23,21) & 1.61 --- 5.00 & 3.39 & W\tablenotemark{g}  & 4.43 --- 5.00 & $34\pm 5$ \\
6 & (20,11) & 1.73 --- 4.95 & 3.22 & S & 2.93 --- 3.34 & $29\pm 3$ \\
7 & (13,17) & 0.34 --- 4.95 & 4.61 & Y & 2.53 --- 2.88 & $21\pm 2$ \\
8 & (12,20) & 2.88 --- 4.95 & 2.07 & S & --- & --- \\
9 & (29,19) & 2.47 --- 5.12 & 2.65 & W\tablenotemark{g}  & 2.82 --- 3.28 & $16\pm 3$ \\
10& (30,16) & 3.28 --- 5.06 & 1.78 & N  & --- & --- \\
11 (Fig.~3)& (19,28) & 4.14 --- 5.29 & 1.15 & W\tablenotemark{g}  & 4.54 --- 4.66 & $17\pm 2$ \\
12& (12,15) & 4.20 --- 5.75 & 1.55 & U  & --- & --- \\
13 (Fig.~2)& (15,21) & 3.62 --- 9.37 & 5.75 & Y & 7.88 --- 8.11 & $21\pm 2$ \\
14& (28,20) & 6.56 --- 9.14 & 2.58 & U\tablenotemark{f}  & --- & --- \\
15& (21,12) & 6.15 --- 9.32 & 3.17 & Y & 7.65 --- 8.05 & $33\pm 3$ \\
16& (14,16) & 8.05 --- 9.83 & 1.78 &W\tablenotemark{g} & 9.43 --- 9.72 & $16\pm 3$ \\
17& (29,10) & 8.22 --- --- & --- & Y & 8.68 --- 9.03 & $22 \pm 3$ \\
18& (27,16) & 8.62 --- --- & --- & Y & 8.68 --- 9.03 & $17 \pm 3$ \\
19& (18,11) & 8.85 --- --- & --- & U/S\tablenotemark{h} & 9.83 --- 10.12 & $58 \pm 15$ \\
20& (14,16) & 9.03(?)--- 9.83\tablenotemark{i} & --- & U\tablenotemark{i} &--- & --- \\
21& (23,18) & 8.85 --- --- & --- & U\tablenotemark{f} &--- & --- \\
22& (16,20) & 9.03 --- --- & --- & U\tablenotemark{f} &--- & --- \\
\hline
Averages & --- &  ---  & $2.72\pm1.2$  & --- & --- & $28.0\pm15.5$ \\ 
\enddata
\tablenotetext{a}{Approximate location of event base (in arcseconds) in video f4.}
\tablenotetext{b}{Approximate start and end time (in min) of event in video f4; dashes indicate event started or ended outside the video's time range.}
\tablenotetext{c}{``Y"= clear (apparent) spinning, ``S"=clear splitting without clear spinning, 
``U"=uncertain, ``W''=weak but somewhat uncertain spinning.}
\tablenotetext{d}{Time period in video f4 over which velocity measurements were made.}
\tablenotetext{e}{Estimated velocity of relative twisting or splitting components, along with estimate of uncertainty. 
The last row the gives the mean of the values, along with the (unweighted) 1$\sigma$ uncertainty.}
\tablenotetext{f}{Twisting or splitting not obvious, but could have been missed due to start or end of video.}
\tablenotetext{g}{We suspect weak or short-duration spinning, but uncertain.}
\tablenotetext{h}{Spinning uncertain due to end of movie; splitting prominent.}
\tablenotetext{i}{Mixed up with event 16, so hard to determine start time and motions.}

\end{deluxetable}
\clearpage



\begin{figure}
\epsscale{1.0}
\plotone{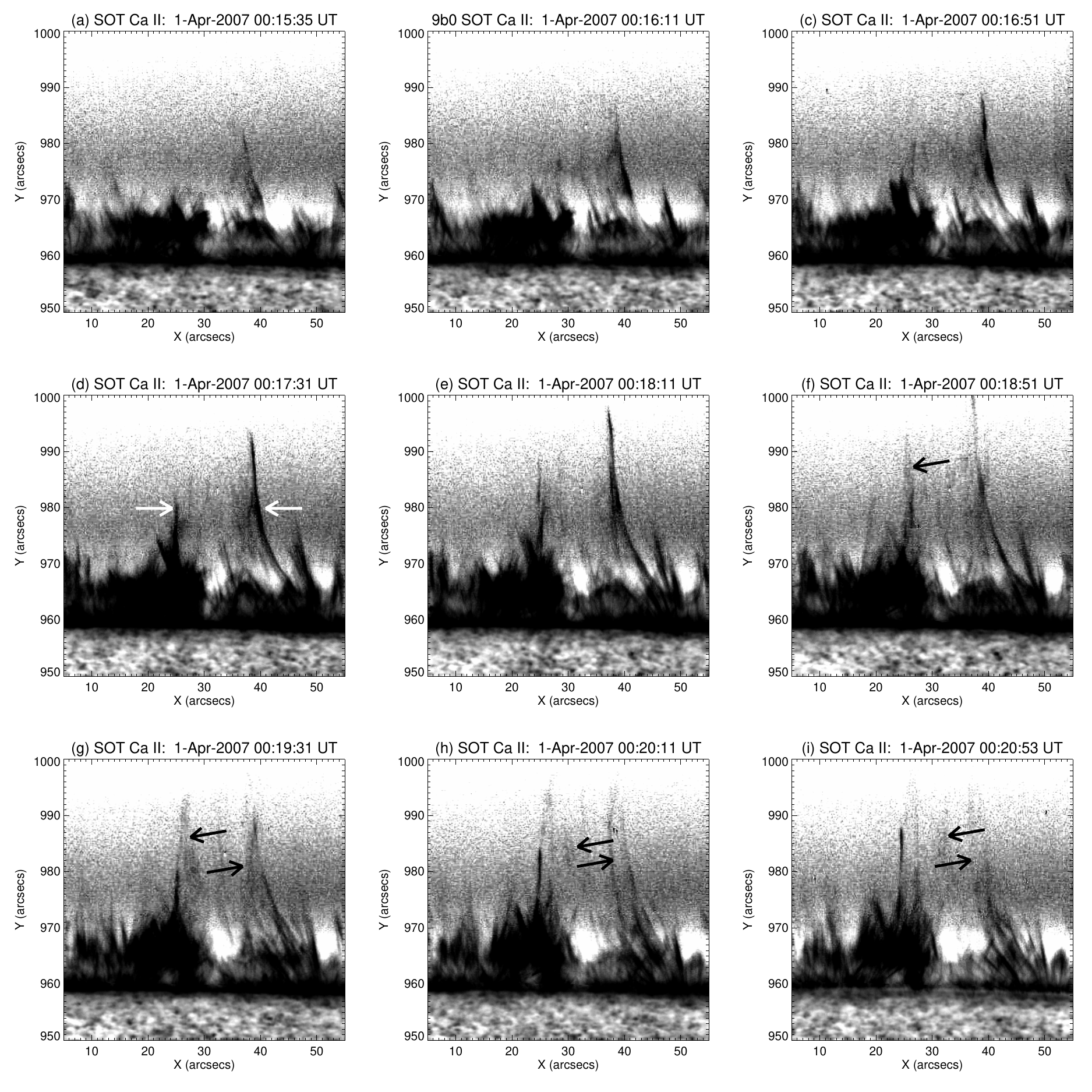}
\caption{Chromospheric component of an X-ray and EUV coronal jet, as discussed by \citet{sterling.et10a}.  These are
\hinode/SOT \caii\ images, where an intensity gradient filter has been applied and colors have been reversed.  The scaling
has been set to show the faint features that appear at the location of the coronal jet, between the two white arrows of (d). 
Black arrows pointing toward the left in (f-i) show a strand of the column  that moves from left-to-right with time, and
similarly the black arrows pointing to the right in (g-i) show a strand moving from right-to-left with time.  These suggest
that we are looking through a partially optically thin cylinder of material, where the cylinder is rotating. Horizontal
intensity strips in the images are an artifact of the radial filter. An accompanying video shows the dynamics of the features with time.  Similar figures and more details are in \citet{sterling.et10a}. North is up and west to the right in this and all other solar images and movies of this paper. This video begins on 2007 April 1, 00:15:35.173 UT and ends the same day at 00:38:14.628 UT. The realtime duration is 3 seconds.}
\end{figure}
\clearpage

\begin{figure}
\includegraphics[angle=270,scale=1.05]{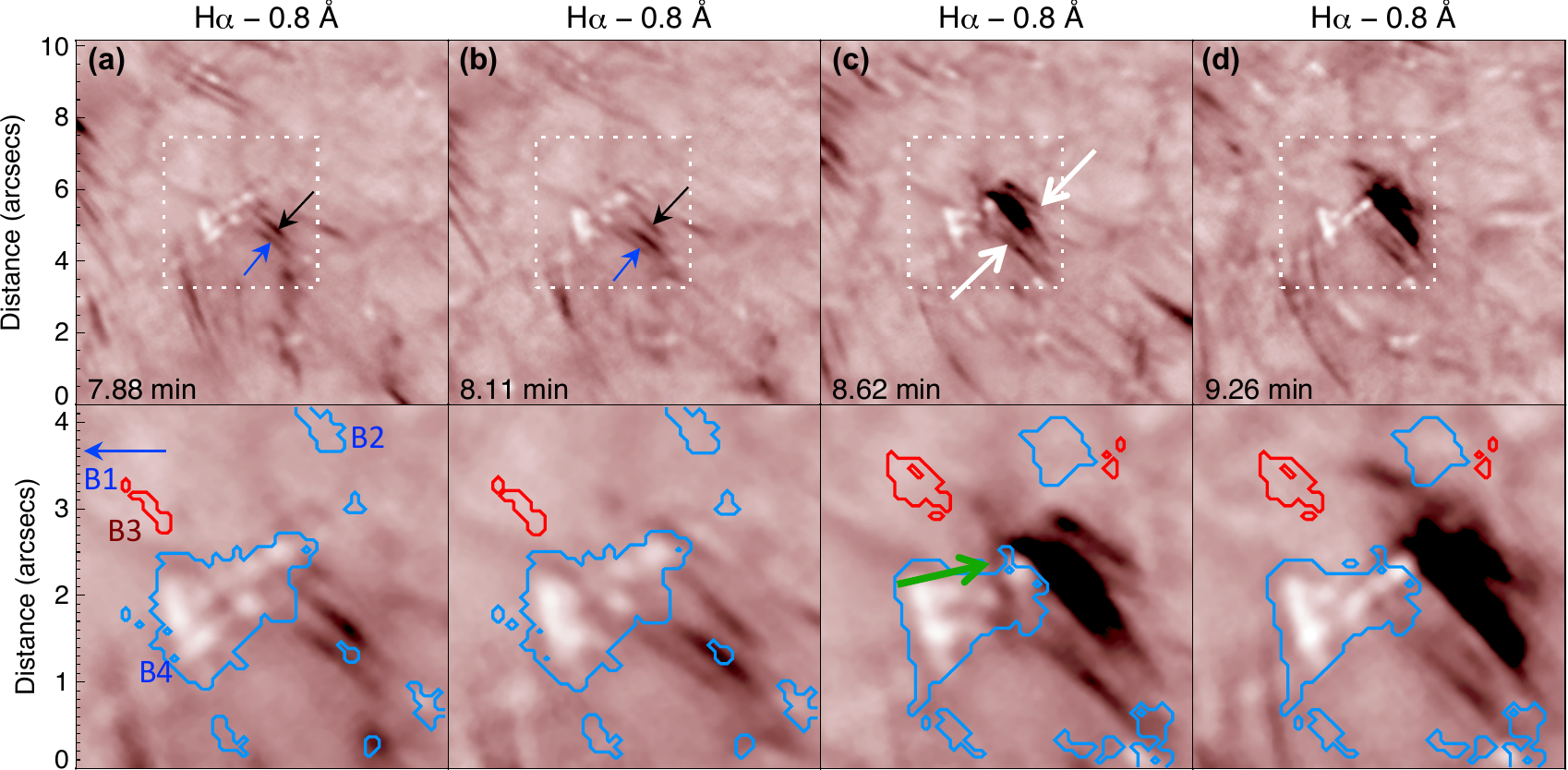}
\caption{Close up of an ``enhanced spicular activity" identified in BBSO/GST observations of
\citet{samanta.et19}.  This is from Fig.~2a of that paper.  For each of (a---c), the top panel is a
$9''.3$ square region, and the bottom panel is a close up of the top panel's dashed white box, and has
size $4''.3$ square.  In (c), the white arrows show the approximate lateral extent (width) of the enhanced spicular activity\@.  In (a-b), the
black arrow points to a strand that moves toward the upper right with time, in a direction opposite to
the that of the strand pointed to by the blue arrows in those panels; these mutual motions suggest that
the enhanced spicular activity feature is undergoing rotation, analogous to that inferred for the
chromospheric manifestation of the coronal jet in Fig.~1.  The green arrow in the bottom panel of (c)
shows an upward-moving horizontally oriented feature that is a candidate for an erupting 
microfilament, which may have been formed by flux cancelation between  negative and positive-polarity
magnetic fields (respectively, red and blue contours in bottom panels), and eruption of which might
produce the enhanced spicular activity. B1-B4 in (a) mark magnetic polarities discussed in Fig.~5.  Fig.~4 shows this event
at the time of 2(c); it is in the southwest of the field of view, near  location (21~Mm,7.5~Mm) in
those panels. An accompanying video shows the dynamics of the features with time. A similar video appears in 
\citet{samanta.et19}.   In these and in other BBSO/GST images and movies, time is measured from the start of the observations, at 18:45:58~UT on 2017 June~19; this video covers from 4.26~min to 10.12~min from that time. The realtime duration of the video is 3 seconds.}
\end{figure}
\clearpage

\begin{figure}
\includegraphics[angle=270,scale=1.3]{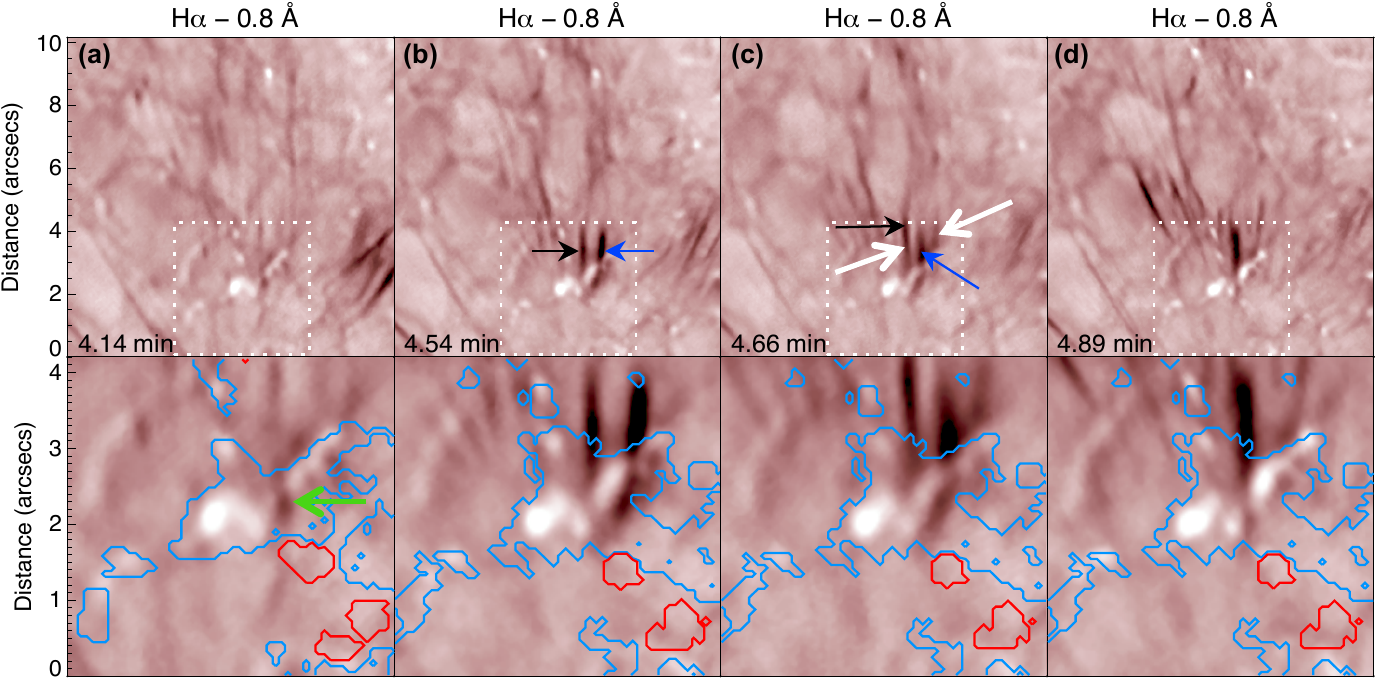}
\caption{Same as Fig.~2, but for the enhanced spicular activity of Fig.~2(c) in \citet{samanta.et19}.  Again the white
arrows in (c) indicate the width of the enhanced spicular activity.  In this case the entire enhanced spicular activity whips upward with
time, and there also appears to be relative rotation of the two strands pointed to by the black and 
blue arrows of panels~(b) and~(c).  In the bottom panel
of (a), the green arrow points to a possible erupting microfilament. An accompanying video shows the dynamics of the features with time.  A similar video appears in 
\citet{samanta.et19}.  Time is measured from the start of the observations, at 18:45:58~UT on 2017
June~19; this video covers from 2.19~min to 7.47~min from that time. Its realtime duration is 3 seconds.}
\end{figure}
\clearpage

\begin{figure}
\hspace*{-0.4cm}\includegraphics[angle=0,scale=0.7]{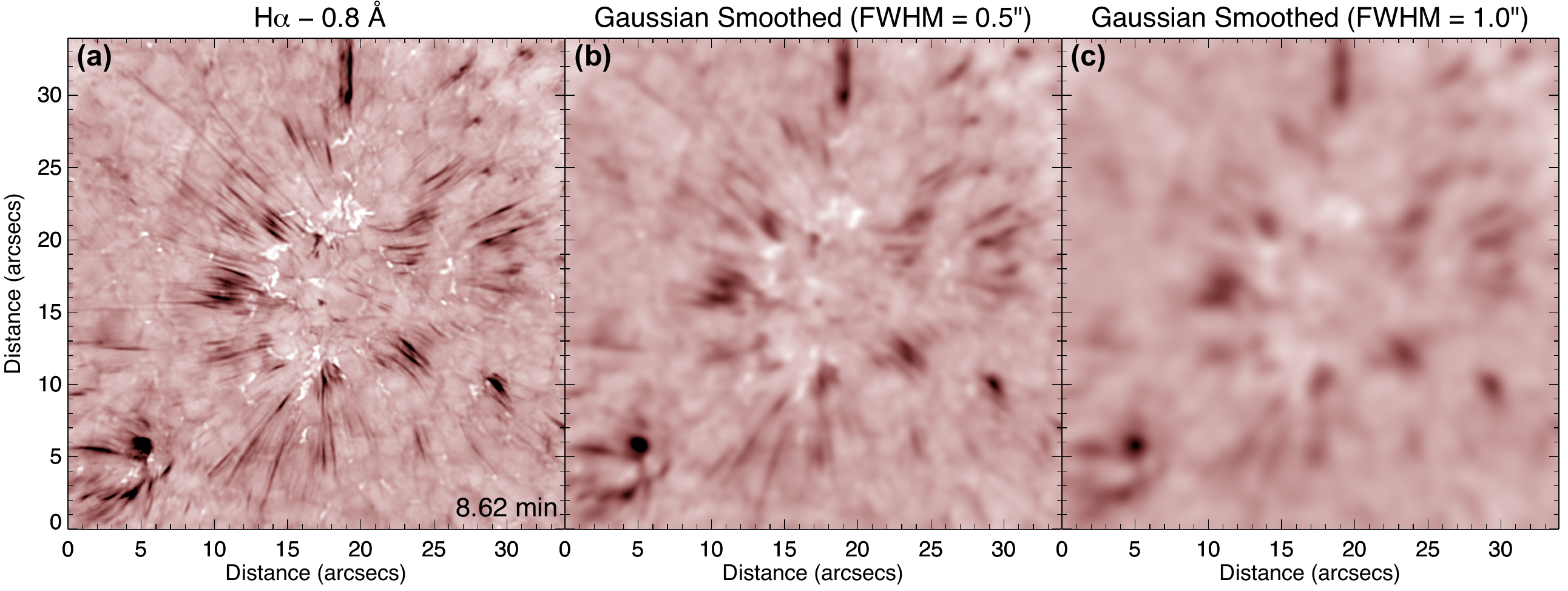}
\caption{The spicule region studied in \citet{samanta.et19}, observed at \halpha-0.8~\AA\ with BBSO/GST, from the time
of Fig.~2(c). Panel~(a) is a 
reproduction of Fig.~1(a) of that paper, but without the magnetic field.  This is shown with the full GST resolution of
$\sim$0$''.06$.  (b) The same images as in (a), but with a Gaussian smoothing with FWHM $0''.5$ applied.  In this case,
details of the enhanced spicular activities are beginning to get lost, as they blur into features of one, two, or a few
linear components; this could explain the double-stranded structure of spicules reported by some observers.  (c) Same as (b),
but with a Gaussian smoothing with FWHM $1''.0$ applied. The enhanced spicular activities now appear to be single blurred
features, similar in appearance to classical spicules/mottles of the Beckers' era. The accompanying video shows the dynamics of the features with time.  A video similar to the left-panel full-resolution video is in 
\citet{samanta.et19}.  Time is measured from the start of the observations, at 18:45:58~UT on 2017
June~19; this video covers from 0~min to 10.12~min from that time, which is the full range of observations
examined in \citet{samanta.et19}.
The realtime duration is 3 seconds.}
\end{figure}
\clearpage

\begin{figure}
\hspace*{0.0cm}\includegraphics[angle=270,scale=0.7]{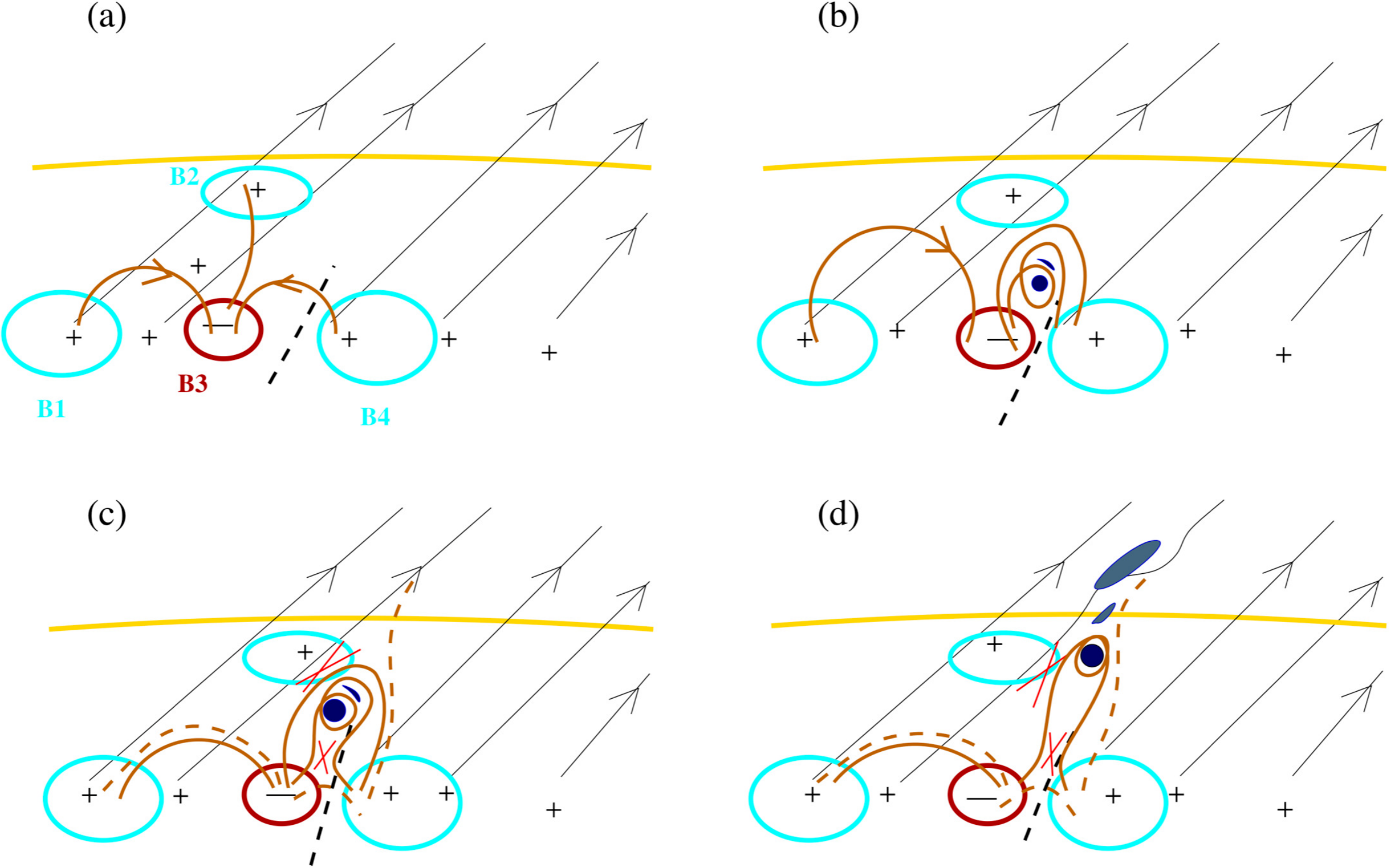}
\caption{Schematic showing the proposed production of an enhanced spicular activity, assuming they are scaled-down
versions of coronal jets that form via eruption of a microfilament flux rope built and
triggered by canceling opposite-polarity 
magnetic fields.  (a) Magnetic setup, tailored to the enhanced spicular activity of Fig.~2(a) viewed from the southeast, where the
yellow curve represents the solar limb.  The blue (positive) and red (negative) magnetic flux patches B1-B4 are as shown 
in Fig.~2(a).  (B1 is not in the displayed field of view of Fig.~2, but a positive field element that could
represent B1 can be seen in Fig.~1 of \citeauthor{samanta.et19}~\citeyear{samanta.et19}.) Plus (+)
and minus (-) signs indicate polarities.  Black solid lines represent open (or far-reaching) magnetic field, and the dashed
black line represents a magnetic neutral line.  Brown lines show the field joining the minority negative polarity element
with surrounding positive field, forming an ``anemone" magnetic structure \citep{shibata.et92}; the field line connecting B3 with B2 in (a)
is omitted from subsequent panels for clarity.   (b) A microfilament 
(dark filled circle) forms in a magnetic flux rope, 
made by the opposite-polarity fluxes converging and canceling at the neutral line.  (c) The microfilament field 
becomes destabilized by further cancelation and erupts, and runaway magnetic
reconnection
occurs at the locations of the red X-es.  Dashed brown lines indicate fields that are newly formed by the reconnections.
(d) As the microfilament eruption continues, external reconnection \citep{sterling.et15} expels some of the 
enclosed microfilament material along the newly open
field lines, forming the strands (dark patches) of the enhanced spicular activity.  Twist that was in the erupting 
microfilament flux rope has been transferred to the far-reaching spicule field, resulting in sometimes-observed (un)twisting motions in spicules.
For right-handed twist in the flux rope, the resulting enhanced spicular activity spins clockwise viewed from above.  This picture is analogous
to the picture of \citet{sterling.et15}, which was proposed to drive coronal jets such as the one 
in Fig.~1.}
\end{figure}
\clearpage

\end{document}